\documentclass[aps,preprint,nopacs,superscriptaddress,12pt]{revtex4-2}
\usepackage{physics}
\usepackage{gensymb}
\usepackage{xspace}
\usepackage{ragged2e}
\usepackage[separate-uncertainty=true,multi-part-units=single]{siunitx}
\usepackage{graphicx}
\usepackage{setspace}

\newcommand{\figtitle}[1]{\textbf{#1}\xspace}

\begin{document}

\setstretch{1}

\title{Programmable, Spontaneous Superlattice Memory in a Monolayer Topological Insulator}

\author{Jian Tang}
\affiliation{Department of Physics, Boston College, Chestnut Hill, MA, USA}

\author{Thomas Siyuan Ding}
\affiliation{Department of Physics, Boston College, Chestnut Hill, MA, USA}

\author{Shuhan Ding}
\affiliation{Department of Chemistry, Emory University, Atlanta, GA, USA}

\author{Jiangxu Li}
\affiliation{Department of Physics and Astronomy, University of Tennessee, Knoxville, TN, USA}
\affiliation{Min H. Kao Department of Electrical Engineering and Computer Science, University of Tennessee, Knoxville, Tennessee, USA}

\author{Changjiang Yi}
\affiliation{Department of Physics, Boston College, Chestnut Hill, MA, USA}
\affiliation{Max Planck Institute for Chemical Physics of Solids, Dresden, Germany}

\author{Tianxing Tang}
\affiliation{Department of Physics, Boston College, Chestnut Hill, MA, USA}

\author{Zumeng Huang}
\affiliation{Department of Physics, Boston College, Chestnut Hill, MA, USA}

\author{Xuehao Wu}
\affiliation{Department of Physics, Columbia University, New York, NY, USA}

\author{Zhiheng Huang}
\affiliation{Department of Physics, Boston College, Chestnut Hill, MA, USA}

\author{Birender Singh}
\affiliation{Department of Physics, Boston College, Chestnut Hill, MA, USA}

\author{Tiema Qian}
\affiliation{Department of Physics and Astronomy and California NanoSystems Institute, University of California Los Angeles, Los Angeles, CA, USA}

\author{Vsevolod Belosevich}
\affiliation{Department of Physics, Boston College, Chestnut Hill, MA, USA}

\author{Mingyang Guo}
\affiliation{Department of Physics, Boston College, Chestnut Hill, MA, USA}

\author{Anyuan Gao}
\affiliation{Department of Chemistry and Chemical Biology, Harvard University, Cambridge, MA, USA}

\author{Nikolai Peshcherenko}
\affiliation{Max Planck Institute for Chemical Physics of Solids, Dresden, Germany}

\author{Zhe Sun}
\affiliation{Department of Physics, Boston College, Chestnut Hill, MA, USA}
\affiliation{Department of Chemistry and Chemical Biology, Harvard University, Cambridge, MA, USA}

\author{Mohamed Shehabeldin}
\affiliation{Department of Physics, Boston College, Chestnut Hill, MA, USA}

\author{Kenji Watanabe}
\affiliation{Research Center for Electronic and Optical Materials, National Institute for Materials Science, 1-1 Namiki, Tsukuba, Japan}

\author{Takashi Taniguchi}
\affiliation{Research Center for Materials Nanoarchitectonics, National Institute for Materials Science,  1-1 Namiki, Tsukuba, Japan}

\author{Abhay N. Pasupathy}
\affiliation{Department of Physics, Columbia University, New York, NY, USA}
\affiliation{Condensed Matter Physics and Materials Science Division, Brookhaven National Laboratory, Upton, NY, USA}

\author{Claudia Felser}
\affiliation{Max Planck Institute for Chemical Physics of Solids, Dresden, Germany}

\author{Kenneth S. Burch}
\affiliation{Department of Physics, Boston College, Chestnut Hill, MA, USA}

\author{Ni Ni}
\affiliation{Department of Physics and Astronomy and California NanoSystems Institute, University of California Los Angeles, Los Angeles, CA, USA}

\author{Yao Wang}
\affiliation{Department of Chemistry, Emory University, Atlanta, GA, USA}

\author{Yang Zhang}
\affiliation{Department of Physics and Astronomy, University of Tennessee, Knoxville, TN, USA}
\affiliation{Min H. Kao Department of Electrical Engineering and Computer Science, University of Tennessee, Knoxville, Tennessee, USA}

\author{Su-Yang Xu}
\affiliation{Department of Chemistry and Chemical Biology, Harvard University, Cambridge, MA, USA}

\author{Qiong Ma}
\email{Corresponding author: maqa@bc.edu}
\affiliation{Department of Physics, Boston College, Chestnut Hill, MA, USA}
\affiliation{The Schiller Institute for Integrated Science and Society, Boston College, Chestnut Hill, MA, USA}

\maketitle
\clearpage

\newpage

\subsection*{Abstract}

\textbf{Memory—the ability to retain information—is a foundational concept across disciplines, from neurobiology and electronics to artificial intelligence and quantum gravity~\cite{keim2019memory}. In materials, memory effects typically arise from ferroic orders, such as ferroelectricity and ferromagnetism, where information is stored in charge or spin degrees of freedom~\cite{auciello1998physics,chikazumi1997physics}. Here, we report a surprising discovery of a nonvolatile \textit{superlattice memory effect} in monolayer TaIrTe$_4$, a dual quantum spin Hall (QSH) insulator~\cite{liu2017van,dong2019observation, guo2020quantum,tang2024dual,lai2024switchable,xi2025terahertz,jiang2025probing}, where information is encoded through sharply contrasting lattice periodicities. In particular, in a \textit{pristine} monolayer, we observe the spontaneous emergence of a long-period superlattice that can be programmed ON and OFF in a nonvolatile manner by electrostatic tuning of low-energy electronic states. This switching toggles the system between two structural configurations with unit cell areas differing by nearly two orders of magnitude. Mechanistically, our results reveal two independent and distinct instabilities—one in the lattice and the other in the QSH electrons—which are coupled, leading to electrostatic control of lattice configurations with nonvolatile memory. This finding is enabled by combining linear and nonlinear transport measurements~\cite{sodemann2015quantum,ma2019observation,kang2019nonlinear,xiao2020berry,du2021nonlinear,gao2023quantum,wang2023quantum2,adak2024tunable,xi2025terahertz,jiang2025probing}, Raman spectroscopy, and scanning tunneling microscopy, which probe complementary aspects of the underlying orders. Remarkably, this nonvolatile memory effect stabilizes a spontaneous superlattice with a periodicity on the few-nanometer scale that remains robust across a wide doping range, persists over days, and survives above 70~K. Combined with the QSH topology, this stability offers a promising route to nonvolatile memory control of topological flat bands and its filling-enabled quantum states. Our preliminary data indeed show the emergence of new insulating states at fractional superlattice fillings, which can be clearly switched ON and OFF together with the superlattice.}\\

Memory plays a central role in both fundamental science and technological applications~\cite{keim2019memory}. In solid-state systems, memory phenomena often originate from spontaneous symmetry breaking. For example, the breaking of spatial inversion and time-reversal symmetries gives rise to ferroelectricity and magnetism~\cite{auciello1998physics,chikazumi1997physics}, respectively (Figs.~\ref{Fig1}\textbf{a-b}), where the associated charge and spin degrees of freedom serve as information carriers. Beyond charge and spin, the lattice configuration itself constitutes a fundamental degree of freedom, governed by translational symmetry breaking and manifested in the (approximate) periodicity of the crystal lattice. This underlying periodicity defines the intrinsic length and energy scales of a material, setting the stage for its electronic structure and phenomena. Recent breakthroughs in two-dimensional moiré superlattices have shown that periodicities on the scale of several nanometers can produce flat electronic bands, enabling a wide range of exotic quantum phenomena~\cite{cao2018unconventional,lu2019superconductors,sharpe2019emergent,serlin2020intrinsic,chen2020tunable,regan2020mott,andrei2021marvels,mak2022semiconductor,wang2022one,kang2023switchable,cai2023signatures,zeng2023thermodynamic,park2023observation,xu2023observation,lu2024fractional,kang2024evidence}—from unconventional superconductivity to fractional quantum anomalous Hall states.\\

Realizing memory control over superlattices—hosts of exotic quantum phases—thus offers a new pathway for manipulating and preserving quantum states, with promising applications in emerging quantum technologies~\cite{wang2017structural,chen2024selective}. However, in conventional systems, the superlattice is typically rigid—either predefined by stack-engineering or dielectric fabrication, as in moiré and dielectric-patterned structures, or fixed during material growth~\cite{forsythe2018band,ribeiro2018twistable,jessen2019lithographic,inbar2023quantum,devarakonda2024evidence,tang2024chip,kim2024electrostatic,wang2025moire}.\\

In this work, we report the surprising discovery of a nonvolatile ``superlattice memory'' in a pristine monolayer, where the system can be switched between two lattice states with sharply contrasting periodicities (Fig.~\ref{Fig1}\textbf{c}). Remarkably, this rare form of memory can be programmed simply via electrostatic gating. Specifically, in the absence of any predefined superlattice structure—either by fabrication or growth—we observe, upon cooling from room temperature, the spontaneous emergence of a superlattice in monolayer TaIrTe$_4$, a recently identified dual QSH insulator~\cite{tang2024dual}. This superlattice can be reversibly switched ON and OFF by tuning the doping level of low-energy QSH electrons, toggling the system between an atomic lattice (unit cell area $\sim$0.47~nm$^2$) and a superlattice (supercell area $\sim$62~nm$^2$).\\

Our systematic measurements reveal that this memory effect likely arises from two coupled but independently evolving order parameters—one in the lattice and the other in the QSH electronic sector. The lattice order is bistable at low temperatures, giving rise to two memory states with distinct structural configurations that are not directly tunable by electric field effects. In contrast, the low-energy electronic order emerges and vanishes directly under electrostatic gating. Its coupling to the lattice provides the driving force for switching between the two structural states, enabling unprecedented electronic programmability of the lattice configuration, which in turn reshapes the electronic properties.\\

In particular, this nonvolatile memory stabilizes a superlattice that not only exhibits a periodicity comparable to twist-engineered moiré systems, but also remains robust across a wide doping range, persists for days (limited only by measurement duration), and survives temperatures above 70~K. Combined with the QSH topology of TaIrTe$_4$, this opens a new pathway to realizing topological flat bands and filling-enabled correlated states. Indeed, we observe transport signatures of superlattice electronic states, including resistance enhancement at fractional filling of the superlattice bands. These findings raise the intriguing possibility of realizing time-reversal-symmetric fractional topological phases that can be nonvolatilely controlled by lattice memory~\cite{levin2009fractional,maciejko2015fractionalized,neupert2015fractional,stern2016fractional,wu2024time,tang2025quantum}.

\subsection*{Dual quantum spin Hall in monolayer TaIrTe$_4$}

\begin{figure*}
\includegraphics[width=6.2in]{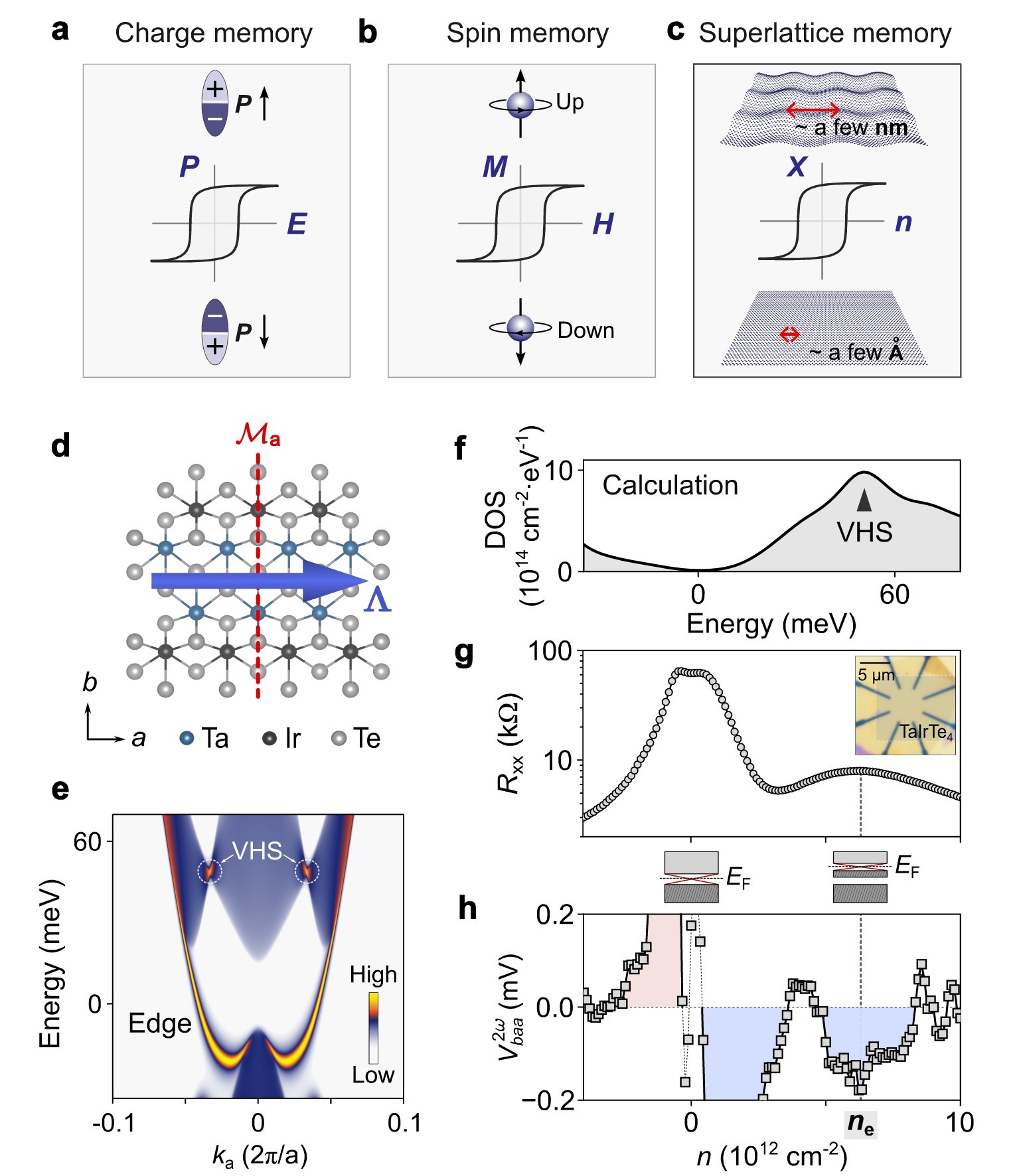}
\caption{\justifying\let\raggedright\justifying{\figtitle{Nonlinear Hall characterization of the dual QSH state in monolayer TaIrTe$_4$.} 
\textbf{a-c,} Illustration of the (\textbf{a}) charge, (\textbf{b}) spin and (\textbf{c}) superlattice memories.
\textbf{d,} Top view of the atomic lattice structure with a mirror plane $\mathcal{M}_{a}$, which permits a Berry curvature dipole $\mathbf{\Lambda}$ along $\hat{a}$.
\textbf{e,} Calculated spectral weight projected along $k_\mathrm{a}$, showing QSH edge states within the bulk band gap and VHSs in the bulk conduction band.
\textbf{f,} Calculated density of state (DOS) versus energy, where a large DOS is observed at the VHSs.
\textbf{g,} The linear resistance $R_{xx}$ as a function of carrier density $n$ at $T=4$ K, where two resistance peaks are observed corresponding to the single-particle band gap ($n = 0$, CNP) and the correlated band gap ($n = n_e$), respectively. Inset shows the optical image of Device 1, scale bar, 5 $\mu$m.
\textbf{h,} The second order ($2\omega$) nonlinear Hall voltage $V^{2\omega}_{baa}$ versus $n$, measured with an AC current $I^{\omega} = 1 \, \mu\mathrm{A}$ at $\omega = 17.777 \, \mathrm{Hz}$ and $T = 4 \, \mathrm{K}$. Large $V^{2\omega}_{baa}$ responses are observed near the CNP and $n_e$.
}}
\label{Fig1}
\end{figure*}

TaIrTe$_4$ is a van der Waals ternary transition metal chalcogenide. Structurally, it consists of one-dimensional, alternating Ta and Ir atomic chains aligned along the crystallographic $\hat{a}$ axis. A mirror plane, denoted $\mathcal{M}_\mathrm{a}$, lies perpendicular to the $\hat{a}$ axis (Fig.~\ref{Fig1}\textbf{d}). Electronically, monolayer TaIrTe$_4$ was predicted to be a QSH insulator with helical edge states (Fig.~\ref{Fig1}\textbf{e}). For a long time, experimental investigation of monolayer TaIrTe$_4$ was limited by fabrication challenges. Only recently has experimental progress enabled access to its intrinsic electronic properties, revealing a novel dual QSH state~\cite{tang2024dual} that extends beyond prior theoretical predictions~\cite{liu2017van,guo2020quantum}. As shown in the experimental work of Ref.~\cite{tang2024dual} and also reproduced in the devices used for this work (Figs.~\ref{Fig1}\textbf{e-g}, Device 1), the dual QSH behavior manifests as follows: at charge neutrality, the system exhibits a QSH insulating state consistent with single-particle band structure calculations. Upon electron doping, the resistance initially decreases and then increases, reaching a second insulating peak at $n_e \approx 6.5 \times 10^{12}$~cm$^{-2}$. This second insulating state, identified as a correlated QSH phase~\cite{tang2024dual}, arises from van Hove singularities (VHSs) in the conduction band (Fig.~\ref{Fig1}\textbf{e}).\\

In contrast to previous studies that relied solely on linear transport~\cite{tang2024dual}, the present work combines linear and nonlinear transport measurements to uncover new phenomena. Specifically, we focus on the second-order nonlinear Hall effect, in which an electrical current at frequency $\omega$ ($I^{\omega}$) generates a second-harmonic transverse Hall voltage ($V^{2\omega}$)~\cite{sodemann2015quantum,ma2019observation,kang2019nonlinear,xiao2020berry,du2021nonlinear,gao2023quantum,wang2023quantum2,adak2024tunable,xi2025terahertz,jiang2025probing}. This effect has been established as a sensitive probe of Fermi surface Berry curvature dipole~\cite{dzsaber2021giant,ortix2021nonlinear,sinha2022berry,huang2023intrinsic,adak2024tunable}, which enables the detection of a hidden superlattice order in this work, as detailed below.\\

The lattice symmetry of TaIrTe$_4$ permits a Berry curvature dipole $\mathbf{\Lambda}$ along the crystallographic $\hat{a}$-axis. According to theoretical predictions~\cite{sodemann2015quantum}, applying $I^\omega$ along $\mathbf{\Lambda}$ (i.e., the $\hat{a}$-axis) generates a second-order Hall voltage $V^{2\omega}_{baa}$ along the $\hat{b}$-axis. This nonlinear Hall response is expected to be enhanced near the QSH gaps due to concentrated Berry curvature hotspots~\cite{lai2024electric}. Consistent with this expectation, our measurements reveal prominent $V^{2\omega}_{baa}$ signals near both QSH gaps (Fig.~\ref{Fig1}\textbf{h}). We further confirm the intrinsic origin of $V^{2\omega}_{baa}$ and its connection to the Berry curvature dipole through a series of systematic experimental validations, as detailed in the \textit{Methods}, Extended Data Fig.~\ref{nonlinear Hall_checklist}, and SI Section 2--5.

\subsection*{Observation of a hidden state beyond the dual quantum spin Hall}

\begin{figure*}
\includegraphics[width=5.6in]{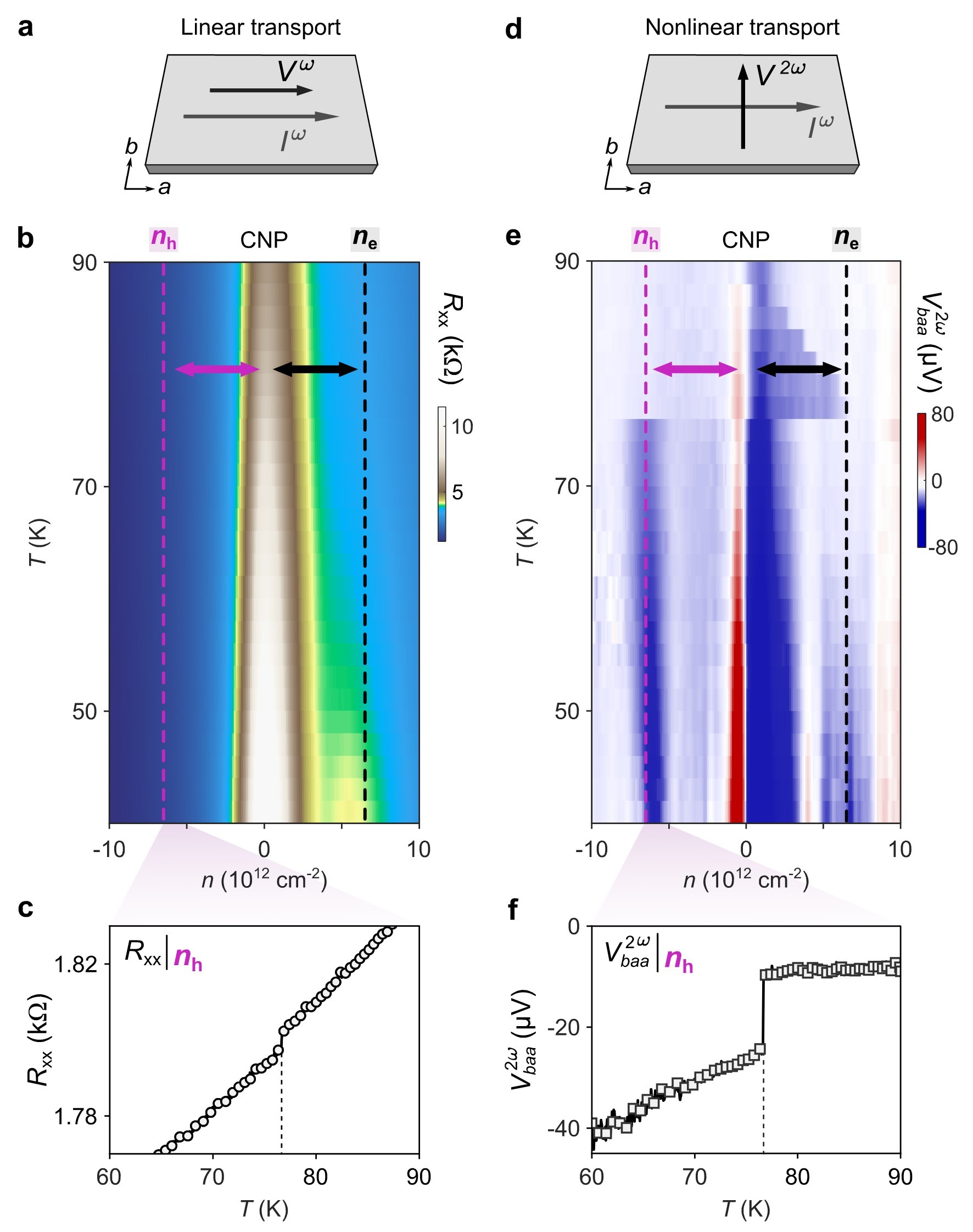}
\caption{\justifying\let\raggedright\justifying
{\figtitle{Observation of a hidden state beyond the dual QSH state.}
\textbf{a–b,} Linear resistance $R_\mathrm{xx}$ as a function of carrier density $n$ and temperature $T$ during warm-up from 40 K. Resistance peaks are observed at the CNP and at $n_e$, corresponding to the single-particle and correlated QSH gaps, respectively. 
As the temperature increases from 4 K to 40 K, the correlated-gap peak shifts to lower carrier density; consequently, $n_e$ appears on the slightly right side of the resistance peak in this panel (Extended Data Fig.~\ref{gap_size}\textbf{a}).
\textbf{c,} Temperature dependence of $R_{xx}$ measured at fixed density $n = n_h$ ($R_{xx}|{n_h}$) during warm-up.
\textbf{d–e,} Nonlinear Hall voltage $V_{baa}^{2\omega}$ measured simultaneously with the data in panel (\textbf{b}).
\textbf{f,} Temperature dependence of $V_{baa}^{2\omega}$ measured simultaneously with the data in panel (\textbf{c}) at $n_h$ ($V_{baa}^{2\omega}|{n_h}$). The detection sensitivity of $V_{baa}^{2\omega}$—defined as the normalized jump across the transition—is 45\%, compared to just 0.13\% in $R_{xx}$, highlighting the enhanced sensitivity of nonlinear Hall measurements in revealing the hidden state. Additional data and analysis are presented in Extended Data Figs.~\ref{Temperature_hysteresis}--\ref{gap_size}.
}}
\label{Fig2}
\end{figure*}

We now compare the linear resistance and nonlinear Hall response over a broader range of carrier densities and as a function of increasing temperature. As shown in Figs.~\ref{Fig2}\textbf{a--b}, $R_{xx}$ exhibits two clear features at $n = 0$ and $n_e$, corresponding to the dual QSH. The simultaneously measured nonlinear Hall voltage, $V^{2\omega}_{baa}$, shown in Figs.~\ref{Fig2}\textbf{d--e}, also captures these dual QSH features at $n = 0$ and $n_e$, consistent with Fig.~\ref{Fig1}\textbf{h}.\\

Strikingly, beyond these, we observe a new feature at $n_h \approx -n_e \approx -6.5 \times 10^{12}$~cm$^{-2}$. This $n_h$ feature emerges only at low temperatures, with an onset around 76~K, indicating the formation of an ordered phase, which we refer to as the \textit{hidden state} (labeled as $V^{2\omega}_{baa}|{n_h}$). We conduct systematic measurements below to uncover its nature. A corresponding discontinuity is also visible in $R_{xx}|{n_h}$ (Fig.~\ref{Fig2}\textbf{c}). However, $V^{2\omega}_{baa}|{n_h}$ exhibits significantly higher sensitivity: the magnitude of the jump—quantified as the normalized difference between pre- and post-transition values—is over two orders of magnitude larger than that observed in $R_{xx}|{n_h}$ (Figs.~\ref{Fig2}\textbf{c,f}). See Extended Data Fig.~\ref{Temperature_hysteresis} for more temperature-dependent data. We also note that although $V^{2\omega}_{baa}\big|{n_h}$ is reliably reproduced across several devices, its behavior below 40~K is not entirely consistent, suggesting device-dependent additional effects (see Extended Data Fig.~\ref{gap_size}). Below, we focus on the consistent behavior above 40~K.

\subsection*{Memory effect of the hidden state}

\begin{figure*}
\includegraphics[width=6.6in]{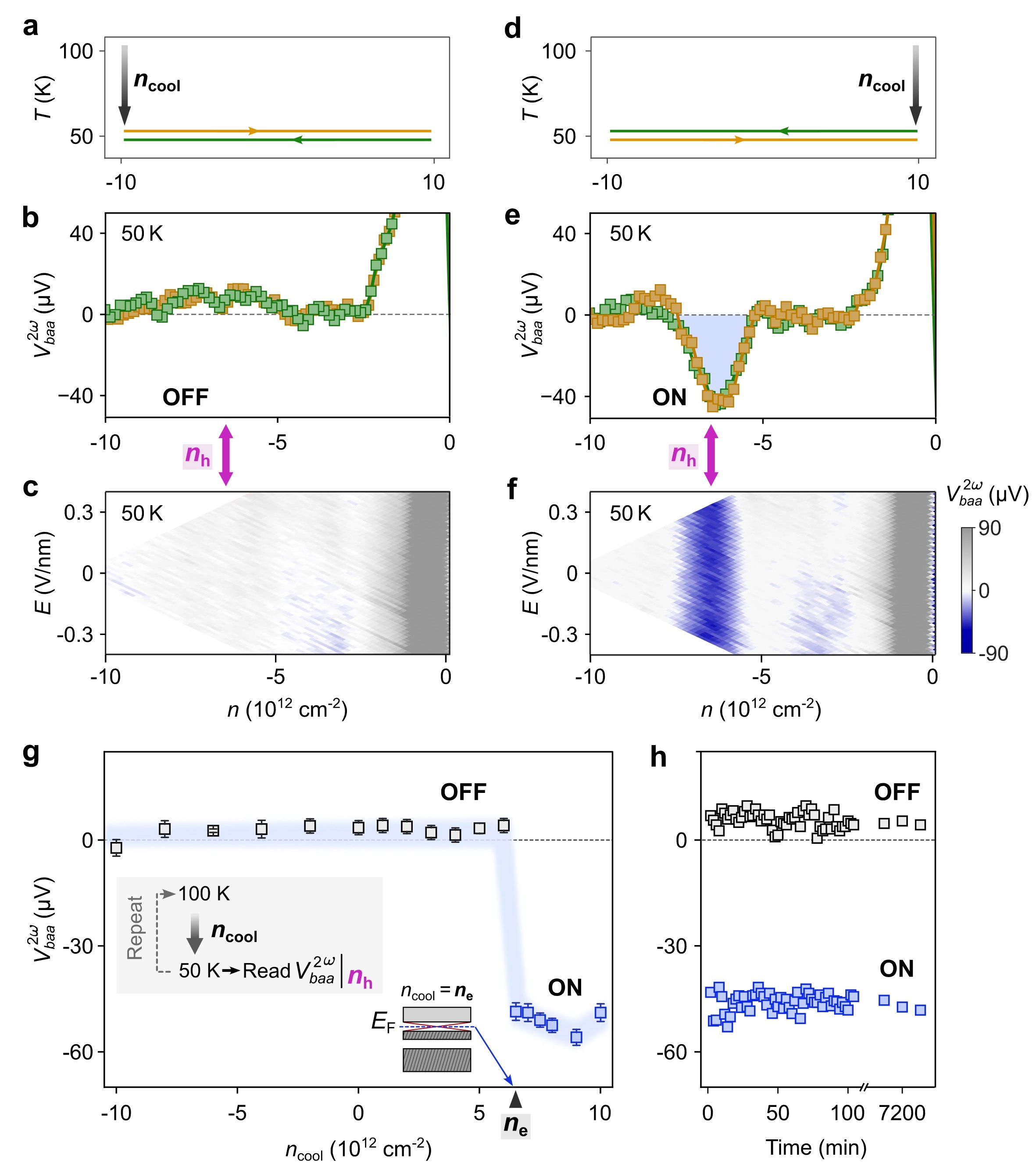}
\caption{\justifying\let\raggedright\justifying{\figtitle{Memory effect of the hidden state revealed by cooling-dependent protocols.}  
\textbf{a-c,} Cooling process 1: The sample is cooled from 100~K to 50~K at a fixed doping level of $n_\mathrm{cool} = -10 \times 10^{12}$~cm$^{-2}$, followed by $V_{baa}^{2\omega}$ measurements at 50~K during both forward (yellow) and backward (green) doping scans. The signal $V_{baa}^{2\omega}|{n_h}$ remains negligible (OFF state) and shows no dependence on the perpendicular electric field $E$.
\textbf{d-f,} Cooling process 2: Same as in (\textbf{a}) but with $n_\mathrm{cool} = +10 \times 10^{12}$~cm$^{-2}$. At 50~K, $V_{baa}^{2\omega}|{n_h}$ exhibits a strong response (ON state), again independent of $E$. 
\textbf{g,} Summary of $V_{baa}^{2\omega}|{n_h}$ responses at 50~K across multiple cooling cycles with varying $n_\mathrm{cool}$. A critical doping threshold, $n_\mathrm{cool} = n_e$, corresponding to the carrier density of the correlated QSH gap, is identified—above which the hidden state is activated.
\textbf{h,} Both ON and OFF states of $V_{baa}^{2\omega}|{n_h}$ persist for several days, limited only by the duration of our measurement.
}}
\label{Fig3}
\end{figure*}

Interestingly, the hidden state is highly sensitive to the doping level during cooldown. We fix the charge density during cooling, denoted as $n_{\textrm{cool}}$, and cool the sample from 100~K to 50~K. After reaching 50~K, we detect the presence of the hidden state by measuring $V^{2\omega}_{baa}$ at $n_h$. The results are shown in Figs.~\ref{Fig3}\textbf{a--f}. When the sample is cooled under a large hole density ($n_{\textrm{cool}} = -10 \times 10^{12}$~cm$^{-2}$, Fig.~\ref{Fig3}\textbf{a}), the hidden state $V^{2\omega}_{baa}|{n_h}$ is absent (Figs.~\ref{Fig3}\textbf{b--c}). In contrast, cooling under a large electron density ($n_{\textrm{cool}} = +10 \times 10^{12}$~cm$^{-2}$, Fig.~\ref{Fig3}\textbf{d}) results in a robust emergence of $V^{2\omega}_{baa}|{n_h}$ (Figs.~\ref{Fig3}\textbf{e--f}). These contrasting outcomes demonstrate that the hidden state is not a simple state function of temperature and doping; rather, it depends on the system’s history—reflecting a memory effect. \\

To determine the critical $n_{\textrm{cool}}$, we perform a systematic study shown in Fig.~\ref{Fig3}\textbf{g}, where $n_{\textrm{cool}}$ is varied stepwise and the presence of the hidden state is evaluated via $V^{2\omega}_{baa}|{n_h}$ at 50~K. The results reveal a critical value of $n_{\textrm{cool}}$, which remarkably coincides with the doping level of the second QSH gap, $n_e$: when the system is cooled under $n \geq n_e$, the hidden state is switched ON; otherwise, it remains OFF. This observation points to a close connection between the hidden state at $n_h$ and the correlated QSH phase at $n_e$, suggesting that the relation $n_h \approx -n_e$ is unlikely to be coincidental. We note that once the system is cooled to 50~K and maintained below this temperature, the presence or absence of the hidden state becomes fixed and unswitchable, exhibiting long retention of at least several days—limited only by the duration of our measurements (Fig.~\ref{Fig3}\textbf{h}).\\

\begin{figure*}
\includegraphics[width=3.6in]{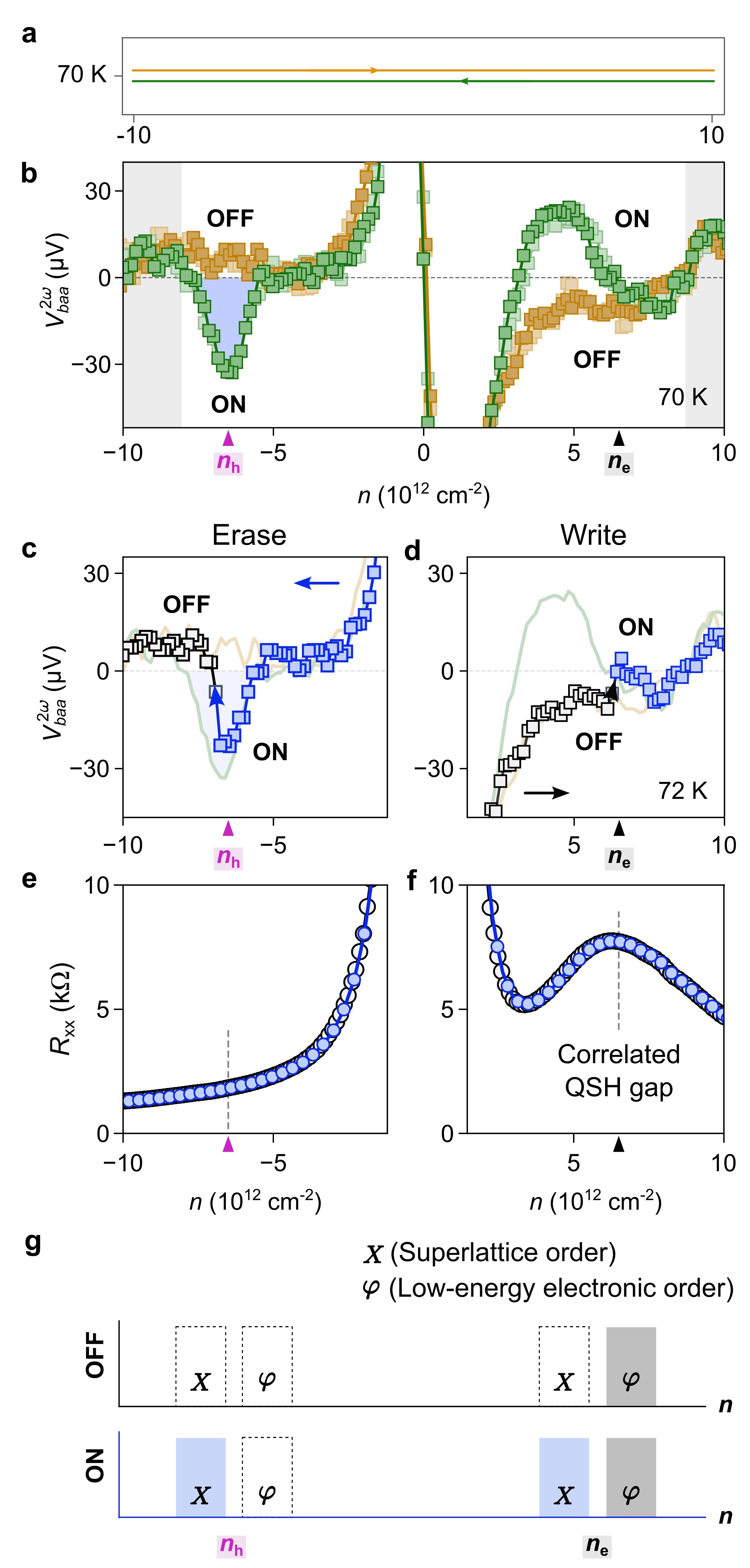}
\caption{\justifying\let\raggedright\justifying
{\figtitle{Direct doping switch of the hidden state and observation of superlattice memory.}
\textbf{a–b,} Forward and backward doping scans of $V_{baa}^{2\omega}$ at $T = 70$~K demonstrate that the hidden state at $n_h$ can be reversibly switched ON and OFF by doping alone. The switching is highly reproducible, as evidenced by two overlapping measurement cycles (green and light green; yellow and light yellow).
\textbf{c–d,} Identification of erase (ON: blue → OFF: black) and write (OFF: black → ON: blue) transitions of the hidden state at $T = 72$~K. Background data from $T = 70$~K (yellow and green) are shown for comparison.
\textbf{e–f,} Comparison of $R_{xx}$ at $T = 4$~K under ON (blue) and OFF (black) states. The correlated QSH gap at $n_e$ remains robust in both cases. 
\textbf{g,} The hidden state corresponds to a superlattice order ($X$) that exhibits ON and OFF configurations at both $n_e$ and $n_h$, whereas the low-energy electronic order ($\varphi$) is always ON at $n_e$ and always OFF at $n_h$.
}}
\label{Fig4}
\end{figure*}

Having established that doping during cooling plays a determining role in the formation of the hidden state, we investigate whether doping can directly toggle the hidden state. We now fix the temperature at $T = 70$~K, which is slightly below the onset temperature of 76~K, and scan the charge density $n$ in both forward and backward directions (Fig.~\ref{Fig4}\textbf{a}). The data reveals a clear hysteresis (Fig.~\ref{Fig4}\textbf{b}): the hidden state is ON when scanning $n$ backward and OFF when scanning forward, thereby directly demonstrating the memory effect. We note that the ON and OFF of the hidden state at $n_h$ are accompanied by a noticeable and highly reproducible change in the features at $n_e$. In fact, Fig.~\ref{Fig4}\textbf{b} shows two complete cycles of forward and backward scans (green and light green, yellow and light yellow), which overlap precisely. This indicates that the order responsible for the response at $n_h$ also exhibits ON and OFF states at $n_e$, a point that will be important for later discussions of the mechanism.\\

The direct doping hysteresis is observed over a temperature range from $\sim$72~K to $\sim$60~K, as systematically shown in Extended Data Fig.~\ref{Cooling+scan_phase}. At each temperature, we can identify a critical ``write" density at which the hidden state switches from OFF to ON. For example, at $T = 72$~K, the transition occurs near $n_e$ during the forward scan, as indicated by the black arrow in Fig.~\ref{Fig4}\textbf{d}. At this point, the OFF-state data (black squares), consistent with the yellow squares in Fig.~\ref{Fig4}\textbf{b}, transition to the ON-state data (blue squares), matching the green squares in Fig.~\ref{Fig4}\textbf{b}. Using a similar method, we can determine the critical ``write'' density at lower temperatures, which will be used to construct the control diagram shown later in Fig.~\ref{Fig6}\textbf{a}. As the system is cooled to lower temperatures, the ``write'' density shifts toward higher electron doping levels. Below 58~K, it becomes impossible to switch the state from OFF to ON within the accessible doping range, indicating that the ``writing force'' is no longer sufficient to drive the transition.\\

Similarly, we seek to identify the carrier density at which the hidden state switches from ON to OFF—the ``erase'' process. As shown in Fig.~\ref{Fig4}\textbf{c} for 72~K, scanning $n$ in the backward direction reveals a jump with a reduced signal area near $n_h$ (blue arrow), indicating a transition from ON to OFF. As the system is cooled to lower temperatures, the ``erase'' threshold shifts toward higher hole doping. However, unlike the ``write'' process, as the ``erase'' density moves away from the $n_h$ dip, the nonlinear Hall response becomes less sensitive—likely because the difference in Berry curvature between the ON and OFF states becomes negligible. As a result, the ``erase'' threshold cannot be determined between 72~K and 61~K. Below 61~K, the state can no longer be switched from ON to OFF within the accessible doping range—the erasing force becomes insufficient to drive the transition. We note that this narrowness of $V^{2\omega}_{baa}\big|{n_h}$ is not universal: in another device, a wider doping-range response allows the erase boundary to be determined more precisely (Extended Data Fig.~\ref{Cooling+scan_phase}).\\

In addition to the nonlinear Hall response, we simultaneously monitor the linear resistance $R_{xx}$ throughout all measurements. We find that the correlated QSH state at $n_e$ is ALWAYS ON, regardless of the hidden state: $R_{xx}$ consistently exhibits a peak at $n_e$, independent of the temperature and doping history (Figs.~\ref{Fig4}\textbf{e--f} and Extended Data Fig.~\ref{linear_2w_ON_OFF}). In contrast, the hidden state—revealed by $V^{2\omega}_{baa}$—exhibits clear ON and OFF behavior that depends on the temperature and doping history, with pronounced signatures not only at $n_h$ but also at $n_e$ (Figs.~\ref{Fig4}\textbf{b} and \textbf{d}). The combined analysis of linear and nonlinear responses is essential for disentangling the nature of the hidden state, as discussed below.

\subsection*{Hidden state is a superlattice – observation of superlattice memory}

We now summarize the key experimental observations and their immediate implications, which form the basis for analyzing the possible origins of the hidden state.\\

\textbf{Observation 1}: We observe a large nonlinear Hall voltage at $n_h$ (Fig.~\ref{Fig2}\textbf{e}), indicating significant Berry curvature near this specific doping level. This response emerges only below an abrupt onset temperature of 76~K, suggesting it originates from a new order beyond the single-particle band structure. This order modifies the band structure near $n_h$, inducing new band hybridization and large Berry curvature.\\

\textbf{Observation 2}: We observe that the doping level $n_h$ is equal in magnitude but opposite in sign to $n_e$, which corresponds to the doping level of the correlated QSH insulator state.\\

\textbf{Observation 3}: We observe that the emergence of the hidden state at $n_h$ is strongly dependent on the doping history. Specifically, Fig.~\ref{Fig3} shows that the hidden state at $n_h$ appears only when the doping level during cooling exceeds a critical threshold near $n_e$. Combined with Observation 2, this highlights a close and intrinsic connection between the hidden state at $n_h$ and the correlated QSH state at $n_e$.\\

\textbf{Observation 4}: We observe that the ON and OFF of the nonlinear Hall response $V_{baa}^{2\omega}|n_h$ are accompanied by a highly reproducible hysteretic change in $V_{baa}^{2\omega}|n_e$, suggesting that the hidden state also influences the electronic properties near $n_e$ (Fig.~\ref{Fig4}\textbf{b}). Meanwhile, the linear resistance $R_{xx}$ consistently shows a peak at $n_e$ (Figs.~\ref{Fig4}\textbf{e--f}), indicating that the correlated QSH insulating state remains robustly ON at $n_e$, regardless of the hidden state.\\

We now examine a few possible interpretations as below:\\

\textbf{Possibility 1}: The nonlinear Hall feature at $n_h$ arises from charge trapping by defects and impurities, which can lead to hysteresis and the memory effect. However, this scenario cannot easily explain the sharp temperature onset behavior and the relation $n_h \approx -n_e$.\\

\textbf{Possibility 2}: The feature at $n_h$ arises from an unintentional superlattice formed between TaIrTe$_4$ and the boron nitride substrate. This scenario naturally explains the relation $n_h \approx -n_e$, as these doping levels correspond to full electron and hole fillings of the same superlattice unit cell. However, in such a case, the value of $n_e$ ($n_h$) would be sample-dependent, varying with the rotational alignment angle. In contrast, our devices consistently exhibit $n_h \approx -n_e \approx 6.5 \times 10^{12}$~cm$^{-2}$. Moreover, this interpretation is not consistent with the observed memory effect or the sharp temperature onset.\\

\textbf{Possibility 3}: The feature at $n_h$ arises from the same order as the correlated QSH state at $n_e$. For instance, the correlated QSH state at $n_e$ may result from a charge order induced by VHSs. This could naturally explain the relation $n_h \approx -n_e$, as both features would originate from the same charge-ordered superlattice. It might also account for the temperature onset and memory effect. However, this scenario is inconsistent with \textbf{Observation 4}. Specifically, the correlated QSH state at $n_e$ is always ON—as confirmed by the persistent peak in $R_{xx}$. If the feature at $n_h$ arose from the same underlying order, it should likewise be always ON at $n_e$. However, experiments clearly reveal two distinct states near $n_e$ in the nonlinear Hall response, which switch in concert with the $n_h$ state (Figs.~\ref{Fig4}\textbf{b} and \textbf{d}).\\

While we do not claim to exhaust all possibilities, we propose a phenomenological framework—our \textbf{Possibility 4}—that naturally accounts for our observations. In this picture, we introduce a low-energy electronic order parameter, $\varphi$, which characterizes the correlated insulating phase at $n_e$. This order, potentially a charge density wave, spin density wave, or another instability that breaks translational symmetry, emerges as doping approaches the VHSs and opens a charge gap at $n_e$. Importantly, this electronic order is always present at $n_e$. In parallel, we introduce a lattice order parameter, $X$, which can take two discrete values: $X \neq 0$ denotes the presence of a superlattice that breaks the same translational symmetry as the electronic order, while $X = 0$ corresponds to its absence. The emergence of the superlattice modifies the band structure near $n_h \approx -n_e$, leading to additional band hybridizations, enhanced Berry curvature, and a pronounced $V^{2\omega}_{baa}$ signal at $n_h$ (see Extended Data Fig.~\ref{ON_OFF_loop}). These features are consistent with \textbf{Observations 1} and \textbf{2}.\\

Furthermore, the presence of the superlattice modifies the Berry curvature near $n_e$, giving rise to distinct features in $V^{2\omega}_{baa}$ at $n_e$ that switch in concert with the ON/OFF transitions at $n_h$, superimposed on the always-ON electronic order at $n_e$—consistent with \textbf{Observation 4}. \textbf{Observation 3} is attributed to the coupling between the electronic and lattice orders, which will be explored in detail in the next section.\\

Therefore, the observed memory effect in $V^{2\omega}_{baa}$ at $n_h$ reflects a memory effect of a spontaneous superlattice, characterized by the lattice order parameter $X$, as illustrated in Fig.~\ref{Fig4}\textbf{g}. The stability of both $X = 0$ and $X \neq 0$ states (Fig.~\ref{Fig3}\textbf{h}) indicates that they correspond to distinct local minima in the free energy landscape, separated by a finite energy barrier.

\subsection*{Superlattice and low-energy electronic orders: independent instabilities and dynamics}

Our experimental observations reveal a distinctive relationship between the superlattice and low-energy electronic orders. Importantly, the superlattice state—reflected by nonlinear Hall responses—can be either ON or OFF at both $n_h$ and $n_e$, whereas the electronic order at $n_e$—evidenced by suppressed conductance—remains robustly ON (Figs.~\ref{Fig3},~\ref{Fig4}). This contrast indicates that the two orders are intrinsically independent instabilities, each exhibiting distinct behavior rather than co-emerging or co-vanishing, as in typical charge density wave or charge-ordered systems.\\

Moreover, we find that the superlattice ON state is triggered when the system is doped to the point where the correlated gap at $n_e$ forms (Fig.~\ref{Fig3}), suggesting a coupling between the lattice and electronic orders. It is natural to conceive that electrical doping—by tuning the Fermi level toward or away from the VHSs—directly controls the electronic instability and its associated order. Through their coupling, the emergence of the electronic order then drives the lattice transition between $X = 0$ and $X \neq 0$ across a finite energy barrier. We can formalize this picture through a Ginzburg–Landau free energy model:

\begin{equation}
\begin{aligned}
    F(\varphi, X; T, n) &= F_{\text{Lattice}}(X) + F_{\text{Electron}}(\varphi) + F_{\text{Coupling}}(\varphi, X), \\
    F_{\text{Lattice}}(X; T) &= \frac{1}{2}\alpha(T) X^2 + \frac{1}{4!}\beta(T) X^4 + \frac{1}{6!} X^6, \\
    F_{\text{Electron}}(\varphi; T, n) &= \frac{1}{2} a(T, n) \varphi^2 + \frac{1}{4!} \varphi^4, \\
    F_{\text{Coupling}}(\varphi, X) &= \lambda \varphi X.
\end{aligned}
\label{F_equation}
\end{equation}

The full analysis of this model and its ability to reproduce the rich control diagram in Fig.~\ref{Fig6}\textbf{a} is presented in the \textit{Methods} and SI Section 5. Here, we highlight several key features. The functional forms of $F_{\text{Electron}}(\varphi)$ and $F_{\text{Lattice}}(X)$ can, even without coupling, exhibit \textit{independent} local minima at nonzero values of $\varphi$ and $X$, as illustrated in Extended Data Figs.~\ref{phase_diagram}. At low temperatures, $F_{\text{Electron}}(\varphi)$ has a single minimum—either at $\varphi = 0$ or $\varphi \neq 0$ (up to sign)—determined by doping. In contrast, $F_{\text{Lattice}}(X)$, modeled by a sixth-order polynomial, supports two coexisting minima at $X = 0$ and $X \neq 0$ (also up to sign), separated by a finite energy barrier. The coupling term $\lambda \varphi X$ enables doping to tilt the free-energy landscape of $X$ and modulate the energy barrier.\\

The energy barrier is essential for decoupling the dynamics of $X$ from $\varphi$ (Extended Data Fig.~\ref{CDW_lattice}): once established, $X$ can remain trapped in either the $X = 0$ or $X \neq 0$ minimum, independent of the current state of $\varphi$. Below, we show how this framework accounts for the key experimental observations in Figs.~\ref{Fig3} and~\ref{Fig4}:\\

\textbf{(1) Doping-induced superlattice switching during cooling:} As shown in Fig.~\ref{Fig3}, cooling under different doping levels leads to either the superlattice OFF ($X = 0$) or ON ($X \neq 0$) state at low temperatures, with a critical threshold near $n = n_e$. This behavior is naturally captured by our free-energy model. As the system cools through the superlattice transition temperature $T_c$—set by $F_{\text{Lattice}}(X)$—the barrier between $X = 0$ and $X \neq 0$ is minimal, allowing even weak writing force $F_\mathrm{write}$ to favor $X \neq 0$. For $n < n_e$, $\varphi = 0$ and the coupling vanishes, keeping the system at $X = 0$. For $n > n_e$, $\varphi \neq 0$ activates the coupling, favoring $X \neq 0$. As cooling proceeds, the barrier grows and locks in the selected state. See Extended Data Fig.~\ref{Free_energy_cooling_ONOFF} for illustrations.\\

\textbf{(2) Doping-induced superlattice switching without cooling:} As shown in Fig.~\ref{Fig4}, sweeping the doping at 72~K directly toggles the superlattice between ON and OFF states with hysteresis. We illustrate the switching cycle using the free-energy landscape in Fig.~\ref{Fig5}\textbf{a}: starting from $n = 0$, the system resides in the $\varphi = 0$, $X = 0$ minimum. As $n$ increases toward $n_e$, $\varphi$ turns on, and the coupling term generates a writing force $F_\mathrm{write}$ that overcomes the energy barrier, switching the system to $X \neq 0$. The required writing density increases at lower temperatures (Fig.~\ref{Fig6}\textbf{a}), indicating that higher electron doping is needed to generate a sufficient writing force to overcome the growing energy barrier.
Upon reversing the doping, $\varphi$ vanishes, but the system remains trapped in the $X \neq 0$ state. The return transition (erasure of the superlattice) occurs near $n_h$, shifting toward higher hole doping at lower temperatures. While the microscopic origin of this return is not fully understood, we discuss possible mechanisms in the \textit{Methods}. Once erased, the system remains in the $X = 0$ state until $F_\mathrm{write}$ again exceeds the barrier, completing the hysteresis loop. See more discussions in the \textit{Methods}.

\subsection*{Raman signature of two lattice states}  

\begin{figure*}
\includegraphics[width=5.5in]{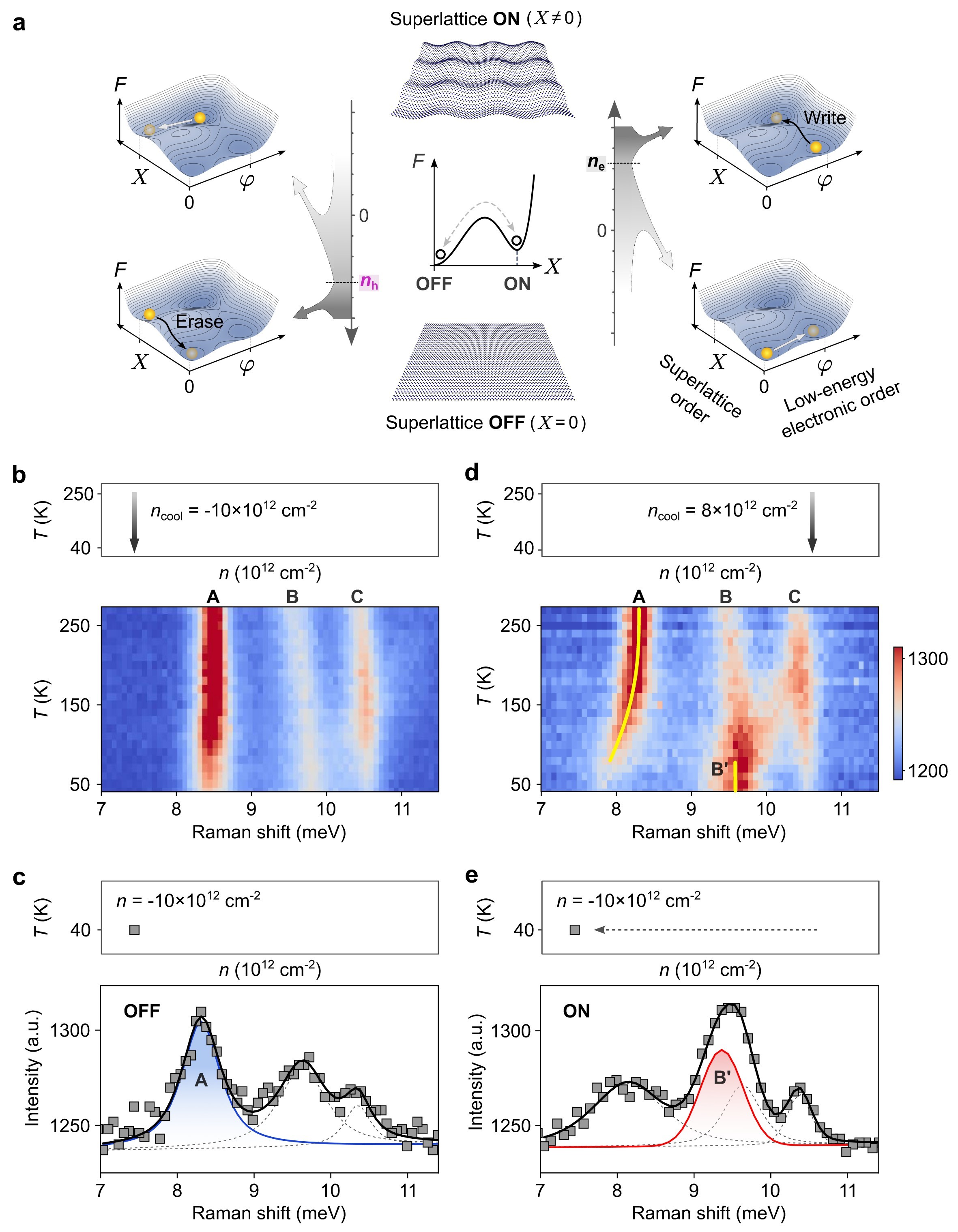}
\caption{\justifying\let\raggedright\justifying{
\figtitle{Electronically driven superlattice and Raman signature of two lattice states.}  
\textbf{a,} Schematic of superlattice switching. Doping tunes the system between $X = 0$ and $X \neq 0$ via coupling to $\varphi$, driving transitions across local minima in the $\varphi$–$X$ free energy landscape. Technically, this occurs through deformation and(or) tilting of the landscape, e.g., as $n$ approaches $n_e$, the minimum shifts from $\varphi = 0$ to $\varphi \neq 0$. We simplify this process with an effective arrow; a more rigorous analysis is provided in SI Section~5. 
\textbf{b-c,} Raman spectra taken while cooling the sample from 270 K to 40 K at $n_\mathrm{cool} = -10 \times 10^{12}$ cm$^{-2}$, corresponding to programming the system into the superlattice ``OFF'' state. (c) shows the spectrum at $T = 40$ K and $n = -10 \times 10^{12}$ cm$^{-2}$.
\textbf{d-e,} Raman spectra taken while cooling the sample from 270 K to 40 K at $n_\mathrm{cool} = 8 \times 10^{12}$ cm$^{-2}$, corresponding to programming the system into the superlattice ``ON'' state. During cooling, peak A softens and a new peak, denoted B$^\prime$, emerges. (e) shows the spectrum at $T = 40$ K after tuning the doping from $n_\mathrm{cool} = 8 \times 10^{12}$ cm$^{-2}$ to $n = -10 \times 10^{12}$ cm$^{-2}$; the B$^\prime$ mode persists.
}}
\label{Fig5}
\end{figure*}

To directly probe the lattice states, we performed Raman spectroscopy using a 532~nm laser with 1~mW excitation. Following the doping-cooling programming procedure, the sample was first cooled to 40~K at $n_\mathrm{cool} = -10 \times 10^{12}$~cm$^{-2}$, which corresponds to the superlattice “OFF” state according to our nonlinear Hall measurements. Under these conditions, the Raman spectrum in the 7--11~meV range exhibits three peaks—labeled A, B, and C—that remain largely unchanged upon cooling; In particular, peak~A remains pronounced, while peak~B stays weak and appears as a single peak feature (Figs.~\ref{Fig5}\textbf{b--c}).\\

In sharp contrast, when the sample was cooled to 40~K with $n_\mathrm{cool} = 8 \times 10^{12}$~cm$^{-2}$, corresponding to the superlattice ``ON'' state (Fig.~\ref{Fig5}\textbf{d}), the Raman spectrum changes dramatically. Upon cooling, peak~A softens by $\sim$0.5~meV, consistent with strong electron--phonon--coupling--induced mode softening, and becomes extremely weak below $\sim$75~K. Simultaneously, an additional peak emerges just below the energy of~B, which we denote as B$^\prime$ (see photon-energy-dependent analysis in Extended Data Fig.~\ref{Raman_red_green}, supporting B$^\prime$ as a new mode). Furthermore, when the system is held at 40~K and the doping is tuned from $8 \times 10^{12}$~cm$^{-2}$ to $-10 \times 10^{12}$~cm$^{-2}$—corresponding to the same doping and temperature as in Fig.~\ref{Fig5}\textbf{c}—the B$^\prime$ mode persists (Fig.~\ref{Fig5}\textbf{e}). These Raman results therefore reveal two distinct lattice states determined by the doping-dependent cooling procedure, further supporting the superlattice memory effect.\\

Interestingly, a recent study~\cite{jiang2025probing} reported CDW transitions in few-layer TaIrTe$_4$ (from bilayer to ten-layer) using nonlinear Hall and Raman spectroscopy. Together with earlier findings in the monolayer~\cite{tang2024dual}, CDW phases appear to be a prevailing feature of the TaIrTe$_4$ family. What is distinctive here is that in our work, both measurements—nonlinear Hall and Raman—consistently reveal two distinct lattice states that can be nonvolatilely switched by tuning the electronic states across the VHSs. In other words, we observe two sets of responses and spectra under identical external conditions (doping and temperature), demonstrating a genuine electronically driven bistability. Such behavior has been lacking in previous studies.

\subsection*{Emergent state at half filling}

This nonvolatile memory effect stabilizes a spontaneous superlattice that remains robust across a wide doping range (see Extended Data Fig.~\ref{CDW_lattice}\textbf{c}), persists for days (limited only by measurement duration), and survives temperatures above 70 K. See discussions of additional performance metrics and future prospects in SI Sections 9, 13 and 14. Moreover, by combining transport data with preliminary scanning tunneling microscopy results, we estimate the supercell area to be $\sim 62\,~\mathrm{nm}^2$ (SI Sections 10-11), comparable to that of flat-band moiré systems.\\

\begin{figure*}
\includegraphics[width=5.4in]{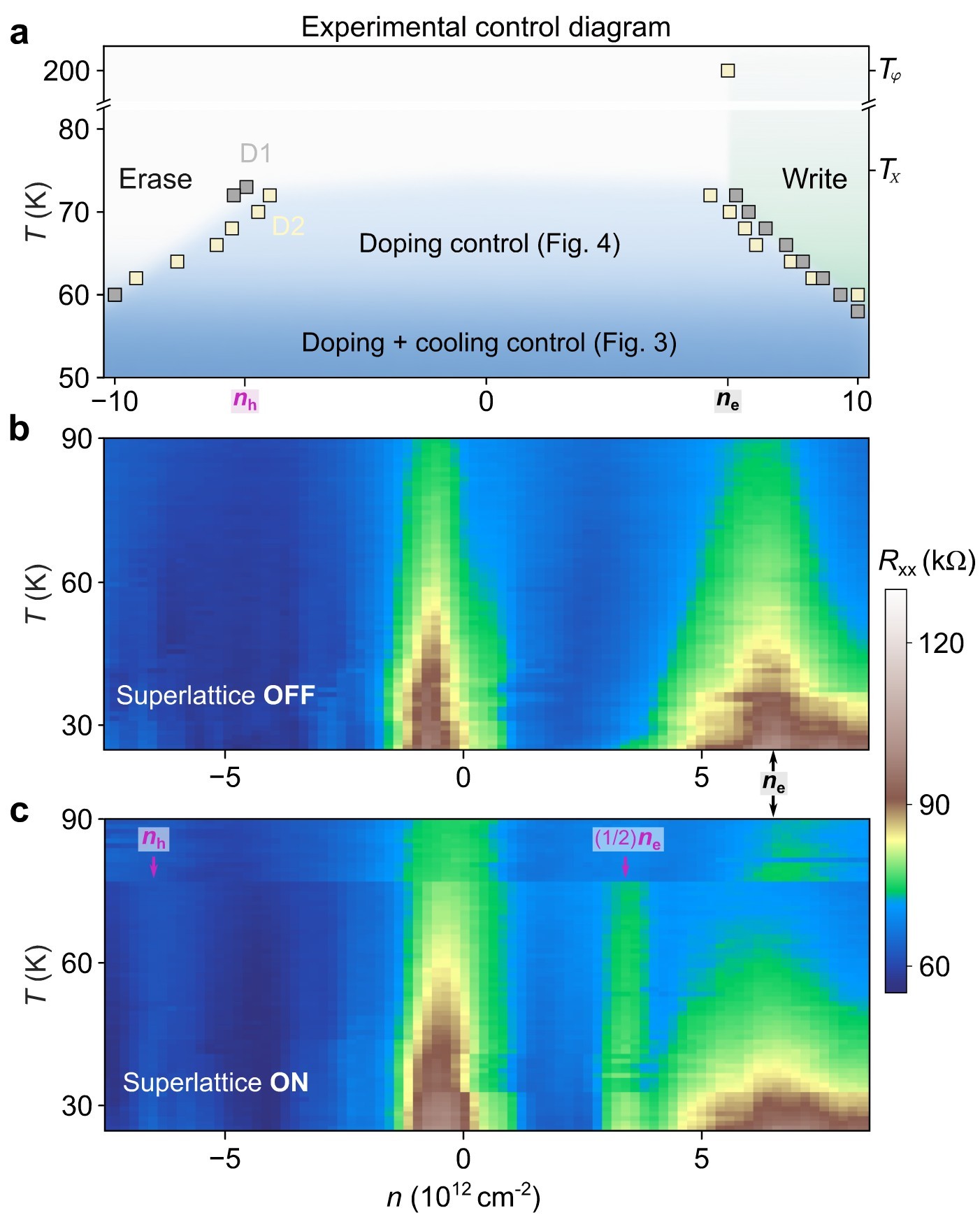}
\caption{\justifying\let\raggedright\justifying{
\figtitle{Experimental control diagram and emergence of fractional superlattice filling states with memory.} 
\textbf{a,} Experimental control diagram. $T_\varphi$ and $T_X$ denote the critical temperatures of the electronic ($\varphi$) and lattice ($X$) orders, respectively. $T_X$ is determined from the temperature dependence of $V_{baa}^{2\omega}$ at $n_h$ (Fig.~\ref{Fig2}\textbf{f}), while $T_\varphi$ is determined from the temperature dependence of $R_{xx}$ at $n_e$ (see SI Section 12). Cooling from the ``Write” region (green) switches the superlattice ON state, whereas cooling from the ``Erase” region (white) sets it OFF state. In the ``Doping + cooling control” region (blue), the state is locked by temperature; in the ``Doping control” region (light blue), it can be reversibly switched by doping alone (Figs.~\ref{Fig3}–\ref{Fig4}). 
\textbf{b-c,} In Device~2, the same programming protocol is applied to set the sample into the superlattice ``OFF'' and ``ON'' states (see SI Section 5 for additional characterizations). The linear resistance $R_{xx}$ maps as a function of carrier density $n$ and temperature $T$ are measured during warming in the OFF state (panel~b) and ON state (panel~c), respectively. The ON state exhibits a weak resistance peak at $n_h$, and an additional distinct peak appears near the filling $\sim\tfrac{1}{2}n_e$, which vanishes sharply around 75~K together with the $n_h$ feature. 
}}
\label{Fig6}
\end{figure*}

Combined with the QSH topology, this platform presents a promising route to realizing exotic correlated phases. In Device~2, we employ the same doping--cooling programming protocol to set the system into the superlattice ``OFF'' (Fig.~\ref{Fig6}\textbf{b}) and ``ON'' (Fig.~\ref{Fig6}\textbf{c}) states. In the ON state, we directly observe a weak but discernible linear resistance peak at $n_h$, further supporting our interpretation of a superlattice-induced gap. In Device~1, this gap is too small to appear in linear transport and can only be resolved through nonlinear Hall measurements. Strikingly, an additional resistance peak emerges near $n_e/2$, marked by the pink arrow. This feature is unique to the ON state and vanishes, together with the $n_h$ peak, abruptly around 75~K. These observations point to a filling-enabled correlated insulating phase within the superlattice flat band. Possibilities include CDW with longer periodicity, Mott, Stoner-type instability, and fractional quantum spin Hall~\cite{levin2009fractional,maciejko2015fractionalized,neupert2015fractional,stern2016fractional,wu2024time,tang2025quantum}. These represent compelling directions for future investigation.

\subsection*{Discussions and outlook}

Beyond the discovery of a nonvolatile superlattice memory and the emergence of spontaneous topological flat bands, we highlight what makes monolayer TaIrTe$_4$ unique—and how it may offer broader implications for other material systems. Electronic and lattice instabilities, along with electron–lattice coupling, are not uncommon in a variety of platforms including kagome metals~\cite{neupert2022charge}, cuprates~\cite{comin2016resonant}, graphene multilayers~\cite{liu2024spontaneous}, and twisted moiré systems~\cite{chen2024strong}.  What distinguishes our system is likely twofold. On one hand, independent electronic and lattice instabilities, with energy barrier and electron--lattice coupling operating at comparable energy scales, are delicately balanced within the free-energy landscape. On the other hand, the monolayer and semimetallic nature enables continuous and selective tuning of the electronic state via doping. Together, these features allow the electronic and lattice orders to be independently accessed, tuned, and stabilized in quasi-equilibrium, with the memory residing exclusively in the lattice sector. This enables new steady-state functionalities and investigation using equilibrium experimental probes.

\newpage

\clearpage
\subsection*{Methods}

\noindent\textbf{TaIrTe$_4$ crystal growth:} High quality of TaIrTe$_4$ crystals were synthesized by a Te flux growth method with a molar ratio of Ta, Ir, and Te elements (purity $> 99.99 \%$) of 3:3:94 in an alumina crucible and sealed in a vacuum quartz tube. The tube was slowly heated up and maintained at 1373 K for 10 hours, and then gradually cooled down to 873 K at a growth rate of 1.5 K/h. The TaIrTe$_4$ single crystals were separated from the Te flux by centrifuging.\\

\noindent\textbf{TaIrTe$_4$ monolayer fabrication:} Monolayer TaIrTe$_4$ was prepared by mechanical exfoliation onto 285~nm-thick SiO$_2$/Si substrates using the Scotch tape method. Due to the air sensitivity of thin TaIrTe$_4$ flakes, exfoliation was conducted entirely inside an argon glovebox. Prior to exfoliation, the SiO$_2$ substrates were treated with mild oxygen plasma to enhance flake adhesion. The tape carrying exfoliated TaIrTe$_4$ was brought into contact with the substrate and gently pressed to eliminate trapped air bubbles. The stack was then heated at $80\,^{\circ}\mathrm{C}$ for 15 minutes. After cooling, the tape was carefully peeled off, and monolayer flakes were identified under an optical microscope.\\

Here, we employed a bottom-contact device architecture with full BN encapsulation. The fabrication began with the assembly of a bottom gate stack (few-layer graphene/BN) using a dry-transfer method facilitated by a PC/PDMS stamp. Residual polymers were removed via thermal annealing at $350\,^{\circ}\mathrm{C}$ for 3 hours in a forming gas atmosphere (Ar/H$_2$) using a tube furnace. Subsequently, metal contacts were patterned using electron-beam lithography (Elionix HS50), followed by deposition of 2~nm Ti/10~nm Pt via e-beam evaporation (Angstrom Engineering). These contacts were then cleaned using both post-annealing and contact-mode tip-cleaning with an AFM. The prepared contact stack was transferred into an argon glovebox for the remainder of the fabrication. Inside the glovebox, we assembled the top gate stack (BN/few-layer graphene/BN) using a PC/PDMS stamp. Monolayer TaIrTe$_4$ flakes were exfoliated, identified, picked up by the top stack, and transferred onto the bottom stack—all within a single day—within the glovebox. Finally, to protect the completed device, a PMMA capping layer was applied via spin coating inside the glovebox. See SI Section 1 for more fabrication details.\\

\noindent\textbf{Linear and nonlinear Hall transport measurements:} The linear ($\omega$) and nonlinear Hall (2$\omega$) transport measurements~\cite{du2021nonlinear,ortix2021nonlinear,adak2024tunable,tang2025quantum} were conducted simultaneously by standard lock-in techniques with a frequency of $\omega$= 17.777 Hz.  Gate voltages were applied using Keithley source meters. In dual-gated devices, the carrier densities $n$ and external electric field $E$ can be independently adjusted through the combination of top gate voltage ($V_\mathrm{tg}$) and bottom gate voltage ($V_\mathrm{bg}$).
\begin{equation}
n=\frac{\epsilon_{\mathrm{0}}\epsilon_{\mathrm{BN}}V_{\mathrm{bg}}}{ed_{\mathrm{b}}}+\frac{\epsilon_{\mathrm{0}}\epsilon_{\mathrm{BN}}V_{\mathrm{tg}}}{ed_{\mathrm{t}}},
\label{n_equation}
\end{equation}
\begin{equation}
E=\frac{1}{2}(\frac{V_{\mathrm{bg}}}{d_{\mathrm{b}}}-\frac{V_{\mathrm{tg}}}{d_{\mathrm{t}}}),
\label{E_equation}
\end{equation}
Here, $\epsilon_{\mathrm{0}}$ is the vacuum permittivity, $\epsilon_{\mathrm{BN}}$ is the BN dielectric constant ($\epsilon_{\mathrm{BN}}\approx 3$) and $d_\mathrm{b}(d_\mathrm{t})$ is the thickness of the bottom (top) BN dielectric layer.\\

\noindent\textbf{Characteristics of nonlinear Hall response:} 
The nonlinear Hall current $\mathbf{J}^\mathrm{NLHE}$ arising from the Berry curvature dipole can be expressed as:
\begin{equation}
\mathbf{J}^\mathrm{NLHE} = \frac{e^3\tau}{2(1+i\omega\tau)}\hat{c} \times \textbf{E}(\mathbf{\Lambda} \cdot \textbf{E}),
\label{NLHE}
\end{equation} 
where $e$ is the electron charge, $\tau$ is the relaxation time, $\omega$ is the frequency, $\mathbf{E}$ is the applied electric field, and $\mathbf{\Lambda}$ denotes the Berry curvature dipole. Based on Equation~\ref{NLHE}, the following key features are expected: (1) \textbf{Hall effect:} The term $\hat{c} \times \mathbf{E}$ indicates that $\mathbf{J}^\mathrm{NLHE}$ is transverse to the applied electric field $\mathbf{E}$, characteristic of a Hall response. (2) \textbf{Angle dependence:} The term $(\mathbf{\Lambda} \cdot \mathbf{E})$ introduces a dependence on the angle between $\mathbf{\Lambda}$ and $\mathbf{E}$. The response is maximized when $\mathbf{\Lambda} \parallel \mathbf{E}$ and vanishes when $\mathbf{\Lambda} \perp \mathbf{E}$. (3) \textbf{Quadratic $I$–$V$:} The nonlinear Hall voltage (current) scales quadratically with the applied current $\mathbf{I}$ (electric field $\mathbf{E}$), regardless of signal polarity. (4) \textbf{Linear scaling between $\sigma_{baa}^{2\omega}$ and $\sigma_{aa}$:} In our case, the second-order nonlinear Hall conductivity $\sigma_{baa}^{2\omega}$ is given by:
\begin{equation}
\sigma_{baa}^{2\omega} = \frac{j_{baa}^{2\omega}}{(E_a)^2} = \frac{e^3\tau}{2\hbar^2(1+i\omega\tau)} \Lambda ,
\label{NLHE_baa}
\end{equation}
With $\omega\tau \ll 1$, this simplifies to 
$\sigma_{baa}^{2\omega} \approx \frac{e^3\tau}{2\hbar^2} \Lambda$,
indicating linear dependence on the relaxation time $\tau$. From the Drude model, the linear conductivity is given by $\sigma_{aa} = \frac{ne^2\tau}{m}$, which is also proportional to $\tau$, suggesting that $\sigma_{baa}^{2\omega} \propto \sigma_{aa}$. These criteria have been systematically examined in our measurements, as shown in Extended Data Fig.~\ref{nonlinear Hall_checklist}, confirming the Berry curvature origin of the observed responses.\\

\noindent\textbf{Raman measurements:} 
Raman spectra were taken in a back-scattering geometry using a home-built setup integrated with an Attodry 2100 cryostat, providing a temperature range from $\sim$1.7 K to 300 K. A dual-gated monolayer TaIrTe$_4$ device was loaded into the cryostat, enabling doping control during Raman measurements. A 532 nm diode laser (Cobalt) or 633 nm He–Ne laser served as the excitation source, with a fixed laser power of 1 mW and an integration time of 200~s per spectrum. Elastic scattering signals were suppressed using a series of notch filters. The Raman setup enables detection of low-energy signals above 20 cm$^{-1}$ with an energy resolution of $\sim$0.5 cm$^{-1}$, using an Andor spectrometer equipped with a 1800 $l$/mm grating. The polarizations are controlled by adjusting the combination of the incident polarizer, quarter-waveplate (QWP) or half-waveplate (HWP) angles, and detection analyzer. The data shown in the main text are measured in the circular cross-polarized configuration controlled by QWP.\\

\noindent\textbf{First-principles calculations:} The Vienna ab initio simulation package (VASP)~\cite{kresse1993ab} has been used to perform the Density Functional Theory (DFT)~\cite{hohenberg1964inhomogeneous} calculations. The valence electron configurations of Ta, Ir, and Te have been described in the frame of projector augmented-wave (PAW) pseudopotentials~\cite{blochl1994projector} while the electronic interactions have been addressed by the Perdew-Burke-Ernzerhof (PBE) functional~\cite{perdew1996generalized}. The cutoff of the plane wave was 500 eV, the width of the Gaussian smearing was 0.05 eV. The $k$-grid of Monkhorst-Pack scheme was set to $16\times6\times1$ for the integrations of Brillouin Zone. For the structure relaxation, the convergence criteria for lattice optimizations were set at $10^{-6}$ eV and 0.1 meV/Å for total energy and ionic forces, respectively. \\

We model the superlattice-modulated band structure using a Fröhlich–Peierls Hamiltonian~\cite{frohlich1954theory} and compute the Berry-curvature dipole, finding qualitative agreement with experimental second-order nonlinear Hall response (see SI Section 3 for details).\\

\noindent\textbf{Free-energy model:} We develop a phenomenological free-energy model with two order parameters (low-energy electronic order and superlattice order) and two external control parameters (doping and temperature). A detailed description of this model and its correspondence with the experiment is provided in SI Sections 6--7. Here, we outline the key points. The free energy $F (\varphi, X; T, n)$ takes the following form:
\begin{equation}
\begin{aligned}
    F(\varphi,X; T, n) = \frac{1}{2!}a(T,n)\varphi^2 + \frac{1}{4!}\varphi^4+ \frac{1}{2!}\alpha(T)X^2 + \frac{1}{4!}\beta(T)X^4 + \frac{1}{6!}X^6 + \lambda\varphi X
\end{aligned}
\label{F_equation}
\end{equation}
Here, $\varphi$ is the low-energy electronic order, $X$ is the superlattice order, $n$ is the carrier density, $T$ is the temperature, and $\lambda$ is the electron-lattice coupling coefficient. Near the phase transition, the coefficients in Eq.~\eqref{F_equation} take the lowest-order temperature ($T$) and carrier density ($n$) dependence as
\begin{align}
    \alpha(T) &= (-c_1T + c_2)^2 + c_3,\\
    \beta(T) &= c_4T - c_5,\\
    a(T, n) &= c_6T - c_7 n + c_8.
\end{align}
We first examine the behavior of individual order parameters without the coupling. The low-energy electronic order $\varphi$-dependence of the free energy $F$ follows the standard Ginzburg-Landau mean-field theory~\cite{mcmillan1975landau}, where a negative quadratic coefficient $a(T,n)<0$ leads to spontaneous symmetry breaking of the electronic order (Extended Data Fig.~\ref{phase_diagram}\textbf{a}). The anharmonicity of the superlattice distortion $X$ is a central feature of our model. Unlike a simple harmonic lattice, where distortions are governed by quadratic terms, the potential landscape of a superlattice distortion requires higher-order anharmonic terms, such as $X^4$ and $X^6$ (Extended Data Fig.~\ref{phase_diagram}\textbf{b}). Anharmonicity generally exists in all crystals and molecules, but it is usually neglected since lattice vibrations are usually confined to the harmonic regime by the steep potential barrier. The anharmonic effects become significantly stronger for superlattices with a long wavelength, which reduces symmetry, folds the phonon spectrum, and leads to a shallower potential surface accessible by electron-phonon coupling.\\

After analyzing the free energy $F$ as a function of individual order parameters, we now examine its behavior in a 2D order-parameter space. In this extended space, $F(\varphi,X)$ develops multiple local minima. To illustrate, we start with the decoupled system, i.e., $\lambda=0$ in Eq.~\eqref{F_equation}. In this limit (Extended Data Fig.~\ref{phase_diagram}\textbf{c}), the local minima of the free energy correspond to the Cartesian product of the one-dimensional minima. At low carrier density far from the VHSs ($n \ll n_e$, Extended Data Fig.~\ref{phase_diagram}\textbf{c} left), the electronic order $\varphi$ vanishes ($\varphi = 0$) for all local minima, and $F(\varphi,X)$ has three local minima, which can be classified based on the superlattice order: we denote the local minima without a superlattice order ($X = 0$) as the type-I phase, while the minima with a finite superlattice order ($X \neq 0$) as the type-II phase.\\

At a carrier density close to the VHSs ($n \geq n_e$, Extended Data Fig.~\ref{phase_diagram}\textbf{c} right), the charge susceptibility is amplified and spontaneous symmetry breaking occurs in the $\varphi$ part of the free energy. The consequence of this ordering in the context of the free energy $F(\varphi,X)$ is the splitting of these potential wells along the $\varphi$ (low-energy electronic order) axis. In this case, with the finite electronic order parameter ($\varphi \neq 0$), we denote the minima without and with superlattice order as the Type-III ($X =0$) and Type-IV ($X \neq 0$) phases, respectively.\\

If the coupling coefficient $\lambda$ is adiabatically increased from zero to its physical value, the local minima of free energy corresponding to phases I, II, III, and IV will be modified accordingly. However, they can still be categorized based on whether they maintain the low-energy electronic order and superlattice order. Note that the electronic order of phase II and the superlattice order of phase III are small but do not exactly vanish due to the finite coupling $\lambda$ (Extended Data Fig.~\ref{phase_diagram}\textbf{d}). The order parameters for the different phases are summarized in Extended Data Fig.~\ref{phase_diagram}\textbf{e}.\\

Extended Data Fig.~\ref{phase_diagram}\textbf{f} shows a phase diagram, featuring four distinct phases. In Phase I, neither the electronic order nor the superlattice is present. This phase resides on the left side of the phase diagram. In Phase II, the electronic order is weak, and the superlattice is ON (denoted as ``1"). This phase occupies the left bottom corner of the diagram. Notably, the left bottom corner can host either Phase I or Phase II, depending on whether the system undergoes $\varphi$-field cooling. Both Phase III and Phase IV correspond to a strong electronic order induced by VHSs and are located on the right half of the diagram. The distinction lies in the superlattice state: Phase III corresponds to superlattice OFF (denoted as ``0"), while Phase IV corresponds to superlattice ON (``1"). The upper part of Phase III occurs because $\varphi$ is not large enough to induce the superlattice. The lower part of Phase III, which overlaps with Phase IV, arises when the cooling path corresponds to no $\varphi$ cooling.\\

In this model, the writing force that drives the transition from $X = 0$ to $X \neq 0$ is clearly the coupling term $\lambda \varphi X$, consistent with experimental observations that the $X \neq 0$ phase is triggered by the electronic order at $n_e$. The reverse transition—from $X \neq 0$ back to $X = 0$—can occur when $\varphi$ vanishes, provided the barrier between the two states is negligible. There is also the possibility of explicit erasing forces arising from coupling between the superlattice and holes. For instance, DFT calculations suggest the presence of VHSs in the valence band below $n_h$~\cite{tang2024dual,lai2024electric,li2025interaction}, which could stabilize an alternative electronic order with a distinct wavevector $Q$ that competes with the order at $n_e$. Such competition may generate an additional erasing force capable of overcoming the barrier and restoring the $X = 0$ state. It would be interesting to explore deeper hole doping toward the valence VHSs, where competing electronic instabilities may stabilize a second superlattice with a distinct periodicity. Observing reversible transitions between the two superlattices would provide strong support for this scenario. We leave this as an open question for future investigation.

\vspace{+5 mm}
\noindent\textbf{Acknowledgements:} 
We thank Meg Shankar, Riccardo Comin, and Kyung-Mo Kim for valuable discussions during the initial Raman measurements. We also thank Justin Song, Liang Fu, Xiaofeng Qian, Yu He, and Ziqiang Wang for insightful discussions. Q.M. and S.-Y.X. acknowledge support from the Center for the Advancement of Topological Semimetals, an Energy Frontier Research Center funded by the US Department of Energy Office of Science, through the Ames Laboratory under contract DE-AC02-07CH11358 (transport measurements). Q.M. acknowledges support from the AFOSR under grants FA9550-22-1-0270 and FA9550-24-1-0117 (sample fabrication and data analysis). Q.M. also acknowledges support from the ONR under grant N00014-24-1-2102 (optical measurements and manuscript preparation), and from the NSF CAREER Award DMR-2143426 (equipment upgrades and maintenance). In addition, Q.M. acknowledges support from the AFOSR DURIP award FA9550-24-1-0077 for the acquisition of equipment used for Raman measurements. Y.Z. acknowledges support from the Max-Planck partner laboratory grant for quantum materials. The research by J.L. was primarily supported by the NSF Materials Research Science and Engineering Center program through the UT Knoxville Center for Advanced Materials and Manufacturing (DMR-2309083). S.D. and Y.W. acknowledge support from the AFOSR Young Investigator Program under grant FA9550-23-1-0153. Simulation results were obtained using the Frontera computing system at the Texas Advanced Computing Center. Frontera is made possible by NSF Award No.~OAC-1818253. Bulk single-crystal growth and characterization of TaIrTe$_4$ were performed at UCLA and were supported by the DOE, Office of Science, under award no. DE-SC0021117. K.S.B. acknowledge support from the AFOSR under award number FA9550-24-1-0110. The work of B.S. was supported by the grant DE-SC0018675 funded by the U.S.Department of Energy, Office of Science. STM work is supported by the Office of Basic Energy Science, Materials Science and Engineering Division, U.S. Department of Energy (DOE) under contractNo. DE-SC0012704. K.W. and T.T. acknowledge support from the JSPS KAKENHI (Grant Numbers 21H05233 and 23H02052) , the CREST (JPMJCR24A5), JST and World Premier International Research Center Initiative (WPI), MEXT, Japan. We also acknowledge that some of the work was carried out in the Boston College cleanroom and nanotechnology facilities. Work by Z.S. was fully completed during his appointment at Boston College, supported from Swiss National Science Foundation under Grant Number P500PT-206914. \\

\noindent\textbf{Author contributions:} JT conducted the experiments under the supervision of QM. JT fabricated the devices with support from TSD, TXT, CY, MG, AG, KSB, and SYX. JT performed the transport measurements and analyzed the data with contributions from TSD, TXT, VB, ZHH, ZMH, ZS, and MS. ZMH built the Raman setup, JT performed the Raman measurements, and analyzed the data with assistance from TSD, BS, ZHH, TXT, and KSB. JL and YZ performed the DFT calculations of the band structures as well as modeling and analysis of the superlattice. SD, NP, YZ, and YW developed the two-order-parameter free energy model and constructed the phase diagram. XW and AP carried out the STM characterizations. CY, TQ, CF and NN grew the bulk TaIrTe$_4$ crystals. KW and TT grew the BN bulk crystals. QM, JT, and SYX co-wrote the manuscript with the input of all authors.\\

\noindent\textbf{Competing interests:} The authors declare no competing interests.\\

\noindent\textbf{Data availability:} Source data are available at
https://doi.org/10.7910/DVN/ULKFAE. All other data that support the plots within this paper and other findings of this study are available from the corresponding authors upon reasonable request.

\clearpage
\setcounter{figure}{0}
\renewcommand{\figurename}{\textbf{Extended Data Figure}}

\begin{figure*}
\includegraphics[width=\textwidth]{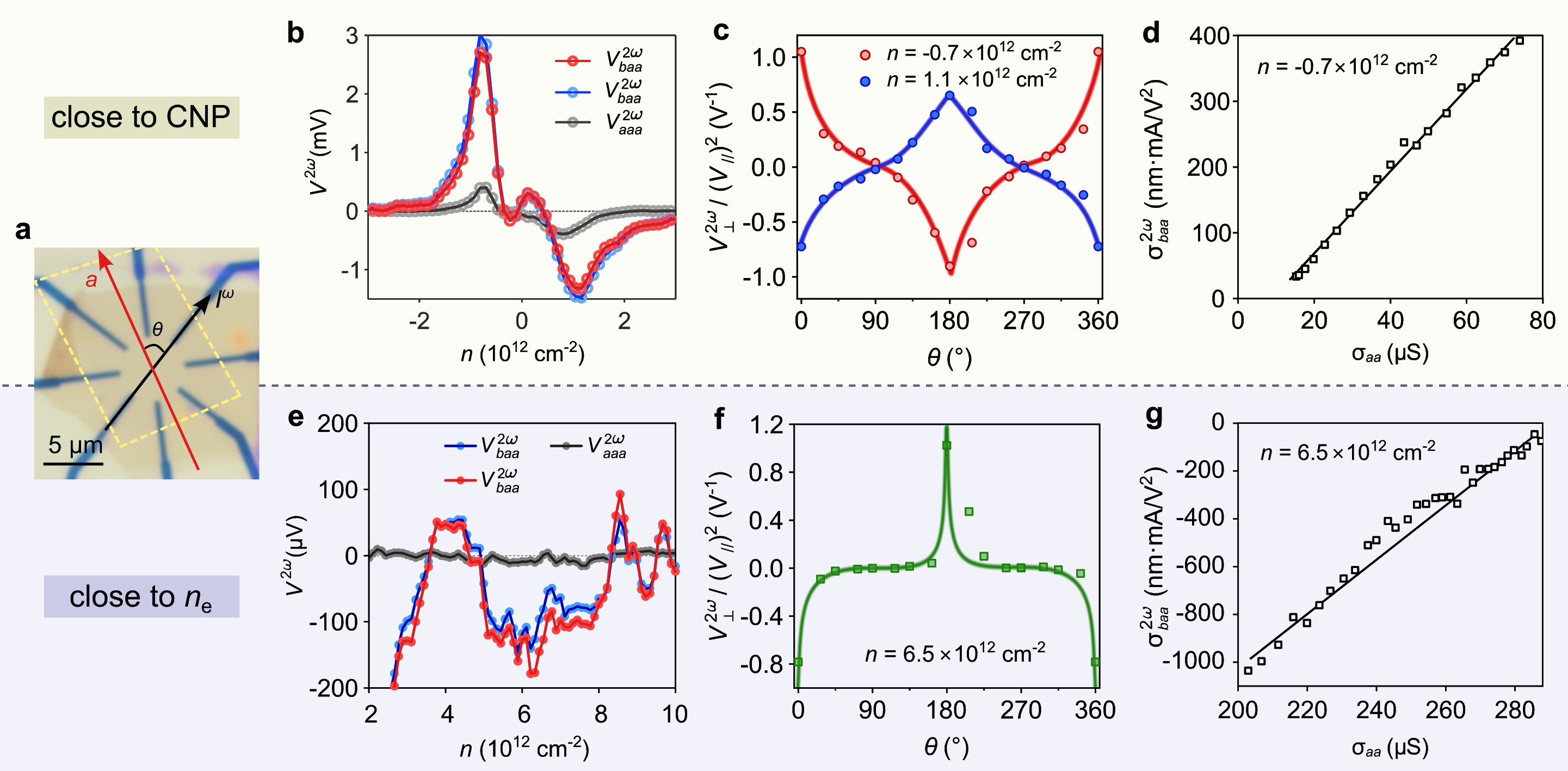}
\caption{\raggedright\justifying{\figtitle{Key characteristics of Berry curvature nonlinear Hall responses.} 
\textbf{a,} Optical image of Device 1, scale bar, 5 $\mu$m. 
\textbf{b-d,} Characteristics of the nonlinear Hall response close to the CNP: (1) The longitudinal $V_{aaa}^{2\omega}$ is significantly smaller than $V_{baa}^{2\omega}$. Note that the two $V_{baa}^{2\omega}$ curves are measured from two parallel sets of Hall probes and exhibit highly consistent results. (2) Angular dependence of the nonlinear Hall response $V^{2\omega}$, consistent with the mirror symmetry $\mathcal{M}_\mathrm{a}$. Here, $\theta$ represents the angle between the crystal axis ($\hat{a}$ direction) and the injection current direction $I^\mathrm{\omega}$, as illustrated in panel \textbf{a}. (3) The scaling relationship between the nonlinear Hall conductivity $\sigma_{baa}^{2\omega}$ and the Drude conductivity $\sigma_{aa}$ is investigated. $\sigma_{baa}^{2\omega}$ can be obtained from: $\sigma_{baa}^{2\omega} \sim \frac{V_{baa}^{2\omega} \, l}{I_a^2 \, R_{aa}^3},$ where $I_a$ is the injected current along $\hat{a}$, $R_{aa}$ is the longitudinal resistance, and $l$ is the sample length, respectively. The observed linear relationship between $\sigma_{baa}^{2\omega}$ and $\sigma_{aa}$ indicates that $\sigma_{baa}^{2\omega}$ depends linearly on $\tau$, consistent with the Berry curvature-induced nonlinear Hall effect.
\textbf{e-g,} Similar characteristics are observed close to $n = n_\mathrm{e}$. Notably, the angular dependence is interestingly sharper.
}}
\label{nonlinear Hall_checklist}
\end{figure*}

\begin{figure*}
\includegraphics[width=6in]{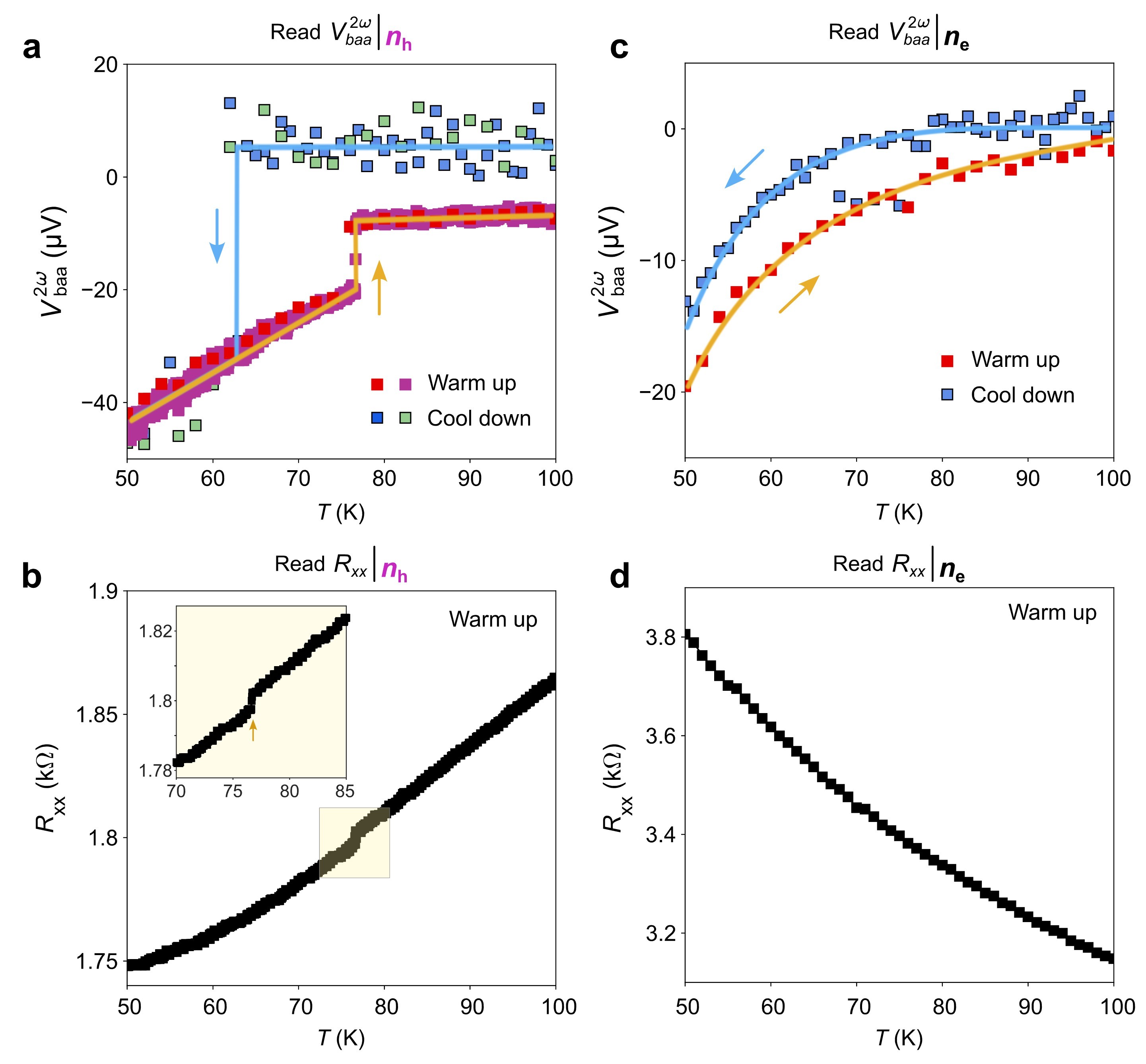}
\caption{\raggedright\justifying{\figtitle{Temperature dependence of nonlinear Hall and linear resistance responses at $n_h$ and $n_e$.} 
\textbf{a,} Temperature dependence of \(V^{2\omega}_{baa}\) at \(n_h\). The cooldown measurement is not straightforward because, to trigger the \(n_h\) response during cooling, the carrier density must first be tuned to \(n_e\) or above. The corresponding measurement protocol and raw data are shown in Extended Data Fig.~\ref{Cooling+scan_phase}\textbf{d} (left). The warm-up measurement is more straightforward, as the system is prepared in the ON state and held at fixed \(n_h\).
\textbf{b,} Temperature dependence of $R_{xx}$, as the system is prepared in the ON state and held at fixed \(n_h\). Inset shows a zoomed-in view of the resistance jump near $T \sim 76$ K.
\textbf{c-d,} Temperature dependence of $V_{baa}^{2\omega}$ and $R_{xx}$ at $n_e$.
}}
\label{Temperature_hysteresis}
\end{figure*}

\begin{figure*}
\includegraphics[width=\textwidth]{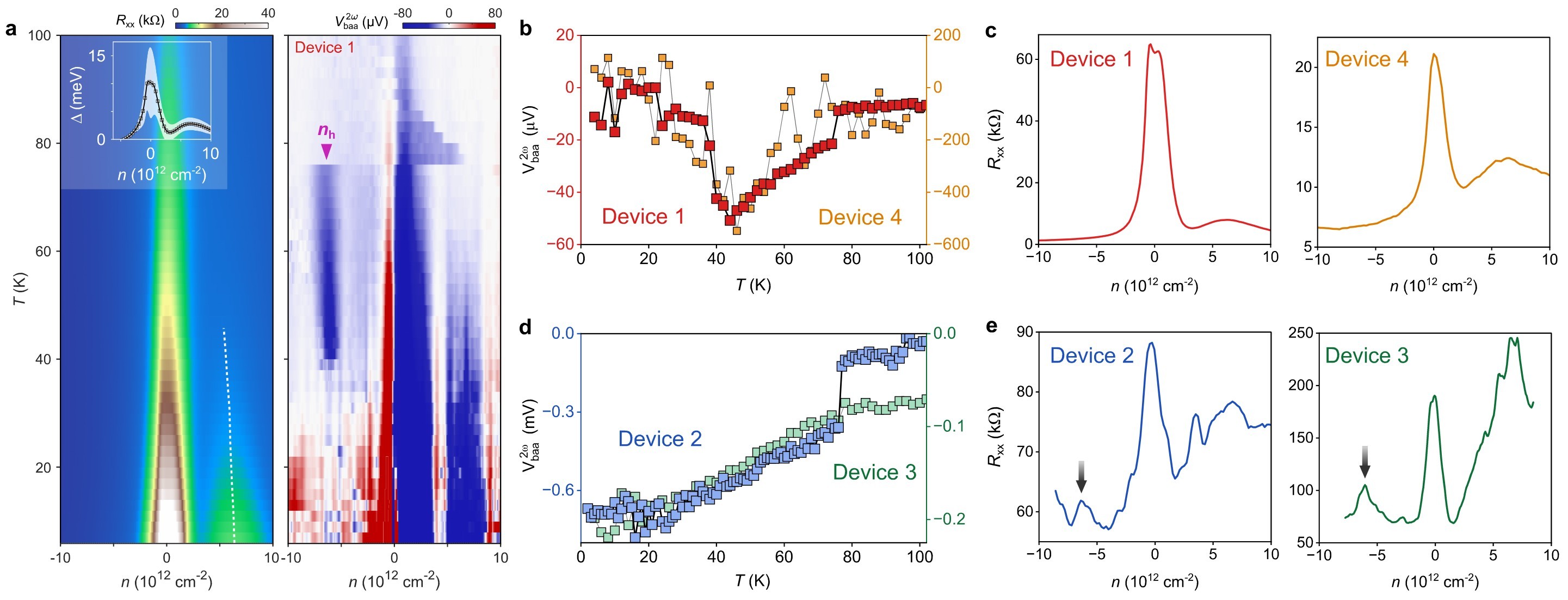}
\caption{\raggedright\justifying{\figtitle{Observations of $R_{xx}$ and $V_{baa}^{2\omega}$ below 40 K.}
\textbf{a,} Simultaneously measured $R_{xx}$ and $V_{baa}^{2\omega}$ as functions of $n$ and $T$ for Device 1, similar to Fig.~\ref{Fig2}\textbf{b} and \textbf{e} but extended to 4 K. The white dotted line labels the \(R_{xx}\) peak near \(n_e\), which shifts to lower carrier density as the temperature increases from 4~K to 40~K.
The extracted gap sizes (inset of the left panel) are smaller than those reported in Ref.~\cite{tang2024dual}, likely due to enhanced inhomogeneity associated with the larger sample size used here, as well as strain effects \cite{li2025interaction}.
\textbf{b--c,} In Devices 1 and 4, $V_{baa}^{2\omega}|{n_h}$ is suppressed below $\sim$40~K, and no $R_{xx}$ peak at $n_h$ (\textbf{c}).
\textbf{d--e,} In Devices 2 and 3, $V_{baa}^{2\omega}|{n_h}$ persists down to 4~K, accompanied by a $R_{xx}$ peak at $n_h$ (\textbf{e}, marked by the black arrows). Based on these empirical observations, together with DFT calculations~\cite{tang2024dual,li2025interaction,lai2024electric} showing hole-side valence-band VHSs at higher doping, we speculate that the suppression of $V_{baa}^{2\omega}|{n_h}$ observed below 40~K reflects a redistribution of Berry curvature driven by correlation effects near the hole-side VHSs.
In Devices 1 and 4, no sizable gap is present at $n=n_h$, making the system more susceptible to such redistribution, with the suppression of $V_{baa}^{2\omega}|n_h$ as a possible consequence. In contrast, in Devices 2 and 3, the presence of a gap at $n=n_h$ may stabilize the electronic structure and Berry curvature against additional correlation effects arising from the hole-side VHSs. Further investigation is needed to fully understand this interesting observation. Note that the gap observed at $n_h$ in Devices 2 and 3 appears only when the device is cooled from densities near the electron-side VHS, consistent with the superlattice mechanism linked to the electron-side VHS.
}}
\label{gap_size}
\end{figure*}

\begin{figure*}
\includegraphics[width=1\textwidth]{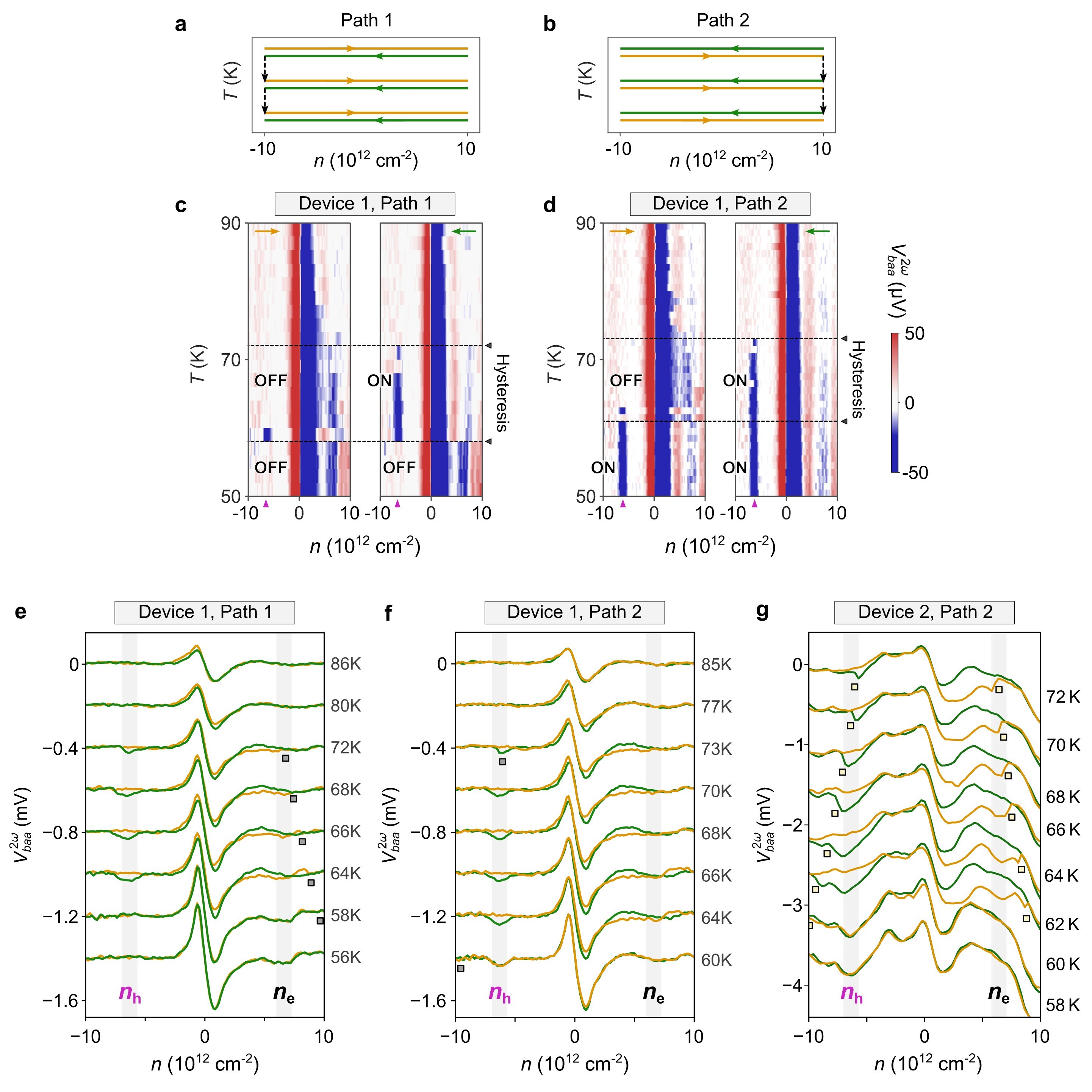}
\caption{\raggedright\justifying{\figtitle{Systematic hysteretic measurements of the hidden state during cooling.} 
\textbf{a-b,} The temperature is stepped down at $n = -10 \times 10^{12}$ cm$^{-2}$ (\textbf{a}, Path 1) and $n = +10 \times 10^{12}$ cm$^{-2}$ (\textbf{b}, Path 2). At each temperature, both forward and backward doping scans are performed to determine the switching points, identified as the densities at which the forward and backward sweeps begin to merge. These points define the critical write boundary for the hidden-state transition from OFF → ON and the erase boundary for the transition from ON → OFF.
\textbf{c-d,} $V_{baa}^{2\omega}$ maps of Device 1 measured following Path 1 and Path 2. A hysteresis window develops upon cooling (72–60 K for Path 1; 73–61 K for Path 2), and the device ultimately stabilizes to distinct final states: $V_{baa}^{2\omega}|{n_h}$ OFF state (\textbf{c}, Path 1); ON state (\textbf{d}, Path 2).
\textbf{e–f,} Hysteresis linecuts from panels \textbf{c–d} for Device 1. The gray square marks the write (OFF→ON) boundary (\textbf{e}, Path 1) and the erase (ON→OFF) boundary (\textbf{f}, path 2).
\textbf{g,} Hysteresis measurement of Device 2 following Path 2, with the write (OFF→ON) and erase (ON→OFF) boundaries indicated by light yellow squares.
}}
\label{Cooling+scan_phase}
\end{figure*}

\begin{figure*}
\includegraphics[width=0.93\textwidth]{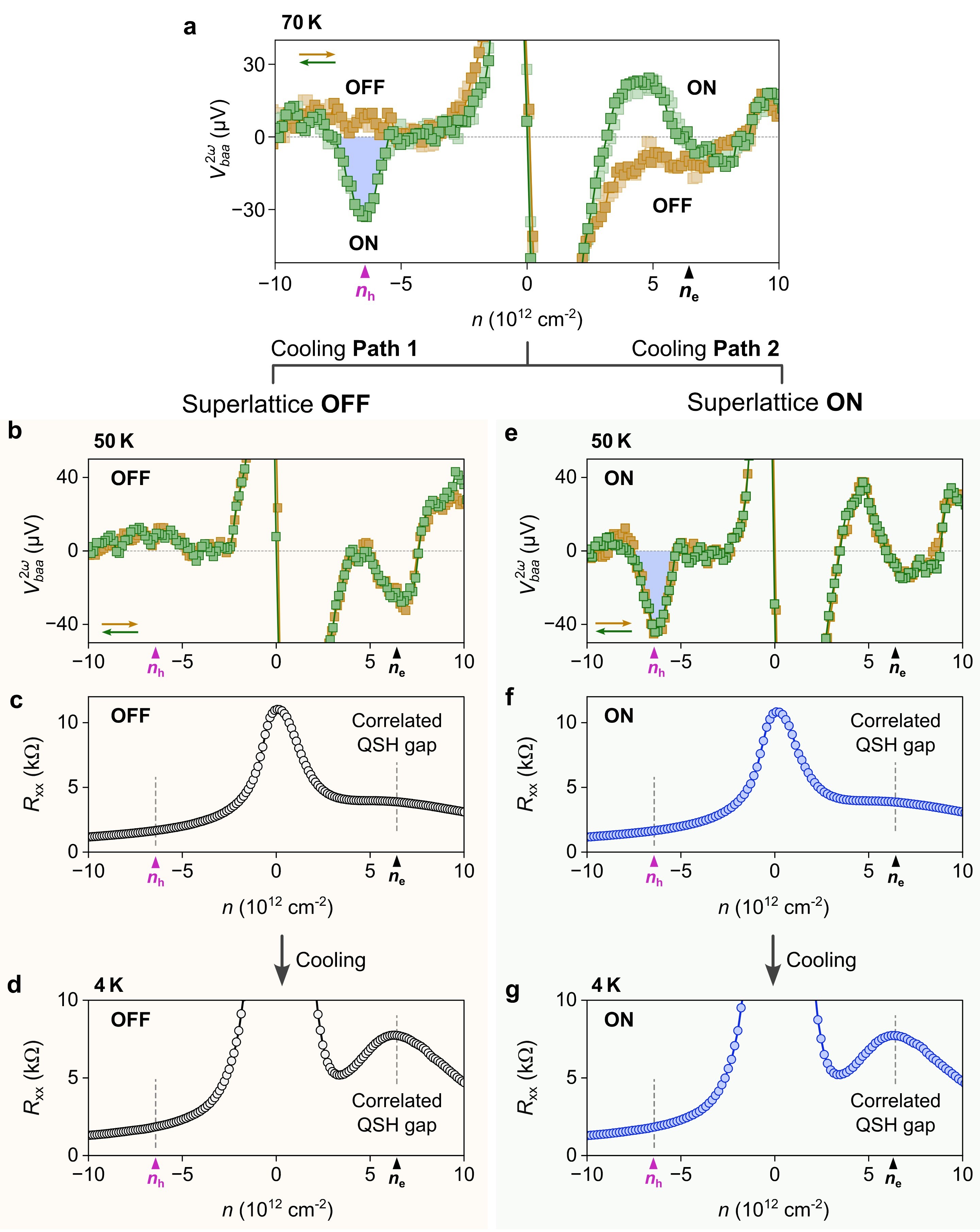}
\caption{\raggedright\justifying{\figtitle{Demonstration of robust ON state of the correlated insulating phase at $n_e$.} 
\textbf{a--d,} Starting from 70~K, the system is cooled along path 1 (cooling with doping away from the VHSs), programming the hidden state to OFF at low temperatures. As shown, $V_{baa}^{2\omega}|{n_h}$ is nearly zero, while the correlated QSH gap at $n_e$ remains robustly ON, as evidenced by $R_{xx}$. 
\textbf{e--g,} Similar to panels \textbf{b--d}, but following cooling path 2 (cooling above the VHSs), which programs the hidden state to ON at low temperatures. In this case, $V_{baa}^{2\omega}|{n_h}$ is large, while the correlated QSH gap at $n_e$ remains ON, as again indicated by $R_{xx}$.
}}
\label{linear_2w_ON_OFF}
\end{figure*}

\begin{figure*}
\includegraphics[width=\textwidth]{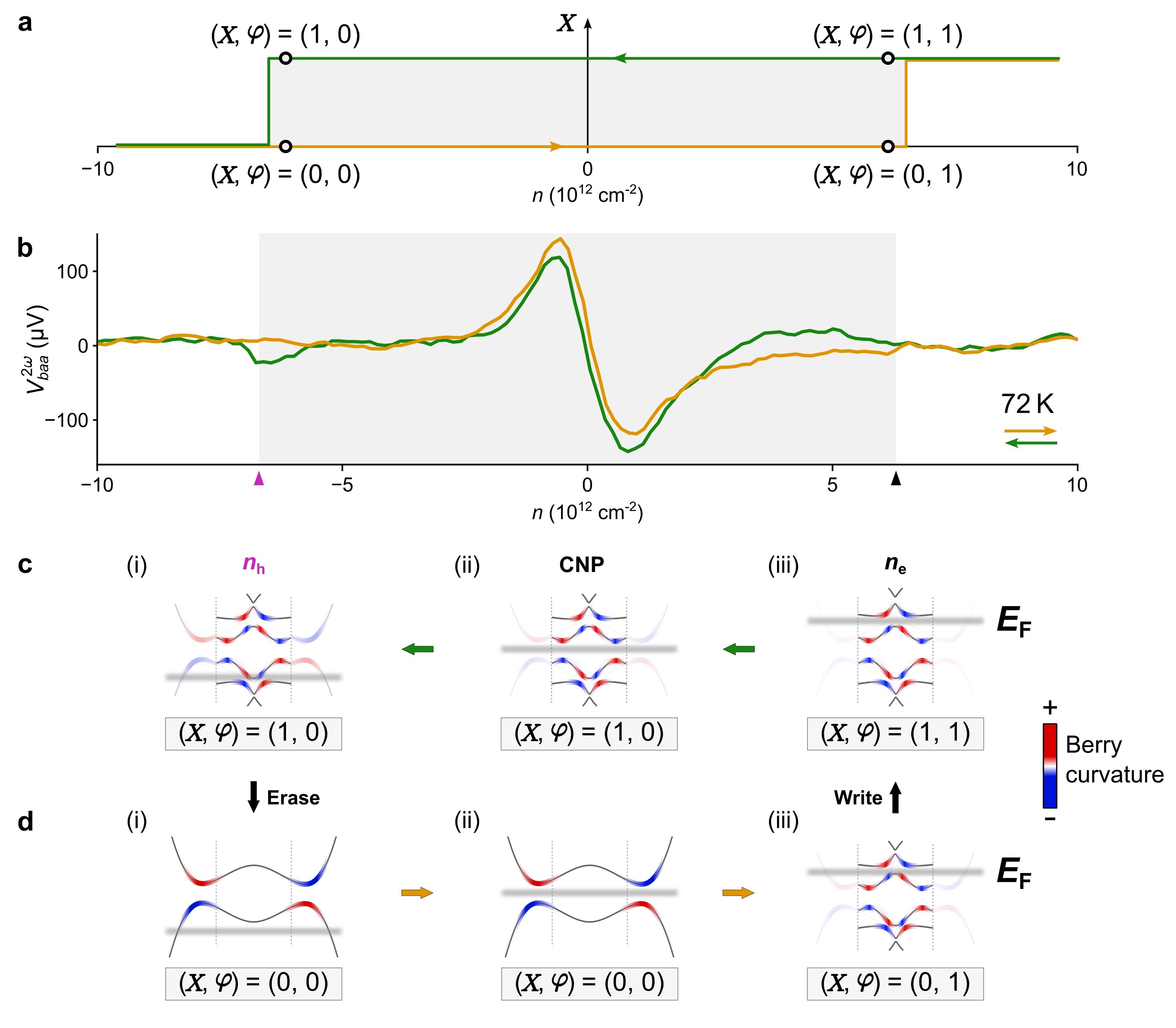}
\caption{\raggedright\justifying{\figtitle{Superlattice ON–OFF contrast in the nonlinear Hall and linear resistance responses at different carrier densities.}
\textbf{a-b,} Schematic illustration of the superlattice memory effect and the associated hysteretic $V_{baa}^{2\omega}$ responses programmed by doping scans.
\textbf{c,} In the superlattice ON state, band folding induced by the superlattice order persists over a wide density range, encompassing both $n_h$ and $n_e$.
\textbf{d,} In the superlattice OFF state, the band structure remains pristine and unfolded at $n_h$ and near the CNP, and becomes folded only when the doping is tuned to $n_e$ due to the electronic order. The two lattice states therefore result in density-dependent ON–OFF contrast. At $n_h$, the OFF state lacks Berry-curvature hot spots, resulting in a nearly zero $V_{baa}^{2\omega}$ response, whereas the ON state exhibits band folding with emergent Berry-curvature hot spots, producing a large $V_{baa}^{2\omega}$ response and thus a strong ON–OFF contrast. Near the CNP, Berry-curvature hot spots already exist in the OFF state due to the topological gap at the CNP; superlattice formation only weakly modifies the band structure and redistributes the Berry curvature, resulting in a small ON–OFF contrast in $V_{baa}^{2\omega}$. Near $n_e$, electronic order alone induces band folding and Berry-curvature hot spots even in the OFF state, while the lattice order provides an additional modulation, yielding a modest ON–OFF contrast in $V_{baa}^{2\omega}$. Similarly, at other carrier densities, switching between the superlattice ON and OFF states modifies the band structure and redistributes the Berry curvature, thereby producing measurable experimental contrast.
}}
\label{ON_OFF_loop}
\end{figure*}

\begin{figure*}
\includegraphics[width=5.3 in]{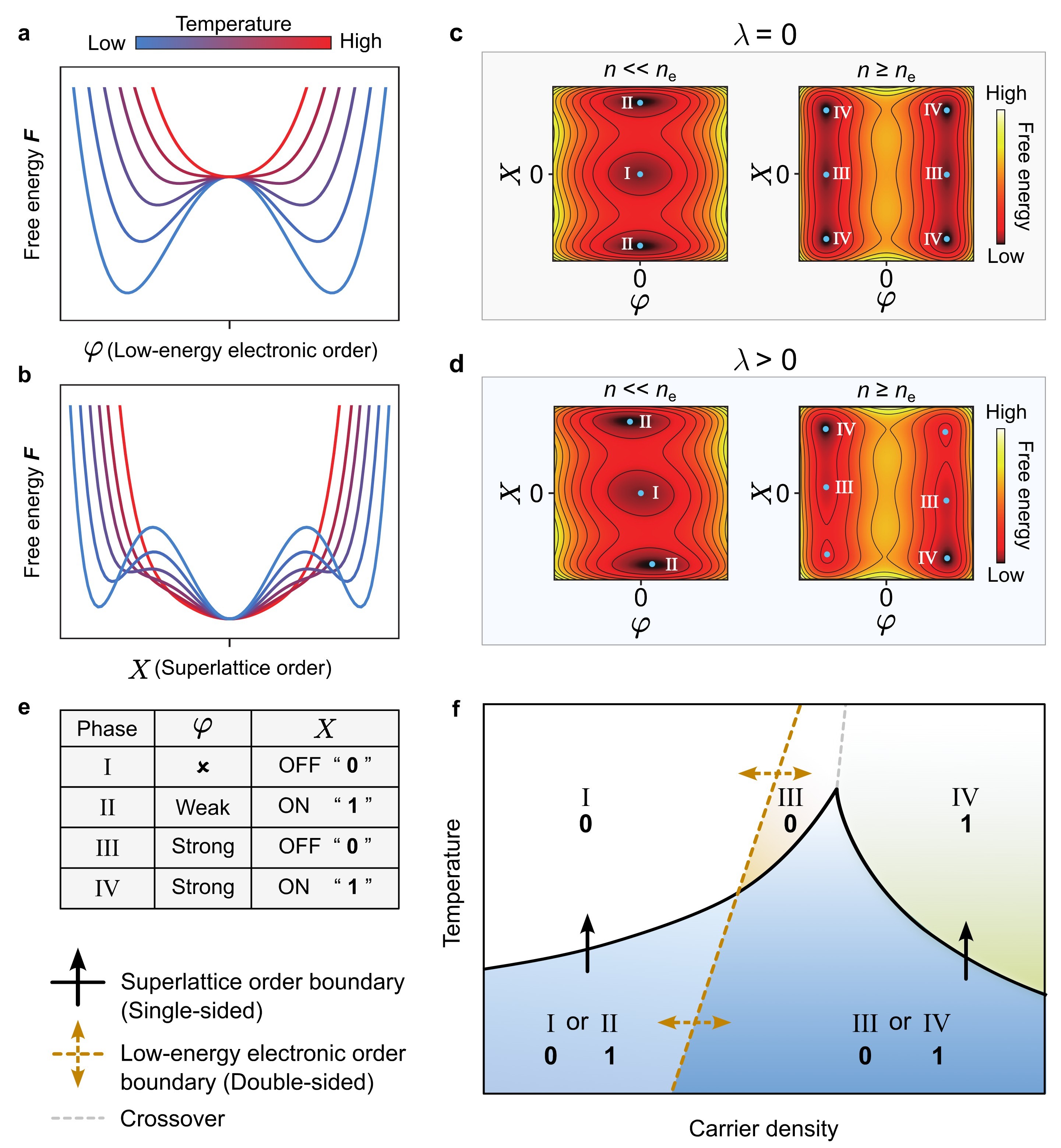}
\caption{\justifying\let\raggedright\justifying{\figtitle{Construction of the phase diagram from the free-energy model.} 
\textbf{a,} Dependence of the free energy $F$ on the low-energy electronic order $\varphi$, with the superlattice order fixed at $X = 0$.
\textbf{b,} Dependence of the free energy $F$ on the superlattice order $X$, with the low-energy electronic order fixed at $\varphi = 0$.
\textbf{c,} In the decoupled regime ($\lambda =0$), the local minima (phases) of the free energy $F$ are classified as Phase I (no electronic order, no superlattice order) and Phase II (no electronic order, with superlattice order) at low carrier density ($n \ll n_e$); and Phase III (with electronic order, no superlattice order) and Phase IV (with electronic order, with superlattice order) at high carrier density ($n \geq n_e$). 
\textbf{d,} Free-energy landscapes at finite coupling $\lambda > 0$.  
\textbf{e,} Summary of the distinct phases determined by the low-energy electronic order $\varphi$ and superlattice order $X$. 
\textbf{f,} Phase diagram: the black solid line represents the phase boundary of the superlattice. Above this boundary, the superlattice order has only one phase (``1" or ``0"); below the boundary, it is bistable, depending on the history. When crossing the boundary from above to below, the superlattice order remains unchanged. However, when crossing from below to above, the superlattice order transitions to the phase above. The yellow dashed line marks the phase boundary of the low-energy electronic order, which is second-order and exhibits no hysteresis. On the left side of this boundary, there is no electronic order or only weak electronic order (due to coupling with the lattice); on the right side, the electronic order is present. The grey dashed line represents a crossover in the superlattice order, rather than a true phase transition.
}}
\label{phase_diagram}
\end{figure*}

\begin{figure*}
\includegraphics[width=6.5 in]{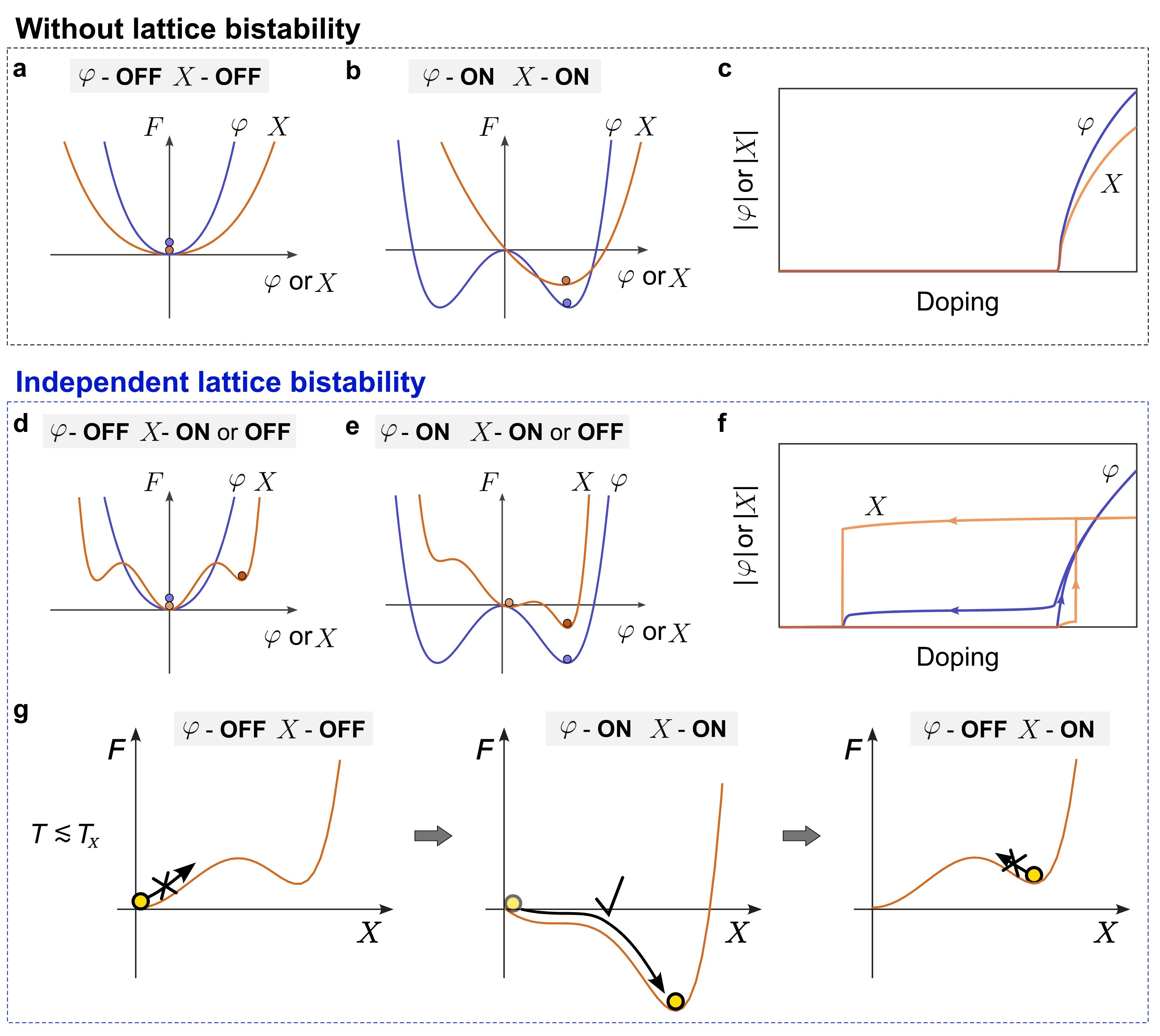}
\caption{\raggedright\justifying
\figtitle{Comparison of free-energy models without and with independent lattice bistability.} 
\textbf{a--c, Without lattice bistability:} We first consider a lattice free energy $F_{\text{Lattice}}$ with a single minimum, modeled as a purely parabolic potential $F_{\text{Lattice}}(X;T)=\frac{1}{2}\alpha(T)X^2$, for which the lattice order is strictly slaved to the electronic order: when $\varphi=0$, $X=0$, and when $\varphi\neq0$, coupling to $\varphi$ shifts the minimum to a finite $X$. Extending $F_{\text{Lattice}}$ to a fourth-order polynomial, $F_{\text{Lattice}}(X;T)=\frac{1}{2}\alpha(T)X^2+\frac{1}{4!}\beta(T)X^4$, yields a unique $X\neq0$ ground state at low temperature, where coupling to $\varphi$ only renormalizes the lattice distortion in a reversible manner. In both cases, no intrinsic bistability or hysteresis is supported, inconsistent with the experimentally observed conditional emergence of the $X\neq0$ state only when $\varphi$ is activated by doping.
\textbf{d--f, Independent lattice bistability:} We next consider the model discussed in the main text, where $F_{\text{Lattice}}$ is described by a sixth-order polynomial supporting two locally stable minima at $X=0$ and $X\neq0$, separated by a finite energy barrier (the sign or phase of $X$ is not distinguished). This intrinsic bistability allows the lattice order to exist in either configuration independently of $\varphi$: both $X=0$ and $X\neq0$ are locally stable for $\varphi=0$ and for $\varphi\neq0$. The coupling term $\lambda\varphi X$ tilts the free-energy landscape, enabling transitions between the two lattice states, while the finite barrier gives rise to pronounced hysteresis.
\textbf{g,} Schematic evolution of the lattice free-energy landscape versus doping. Electron doping induces a finite electronic order $\varphi$, which lowers the $X\neq0$ minimum via $\varphi X$ coupling and enables barrier crossing from $X=0$ into $X\neq0$. Upon removing $\varphi$, the barrier persists, trapping the system in the $X\neq0$ configuration and producing hysteretic, nonvolatile lattice behavior.
}
\label{CDW_lattice}
\end{figure*}

\begin{figure*}
\includegraphics[width=5 in]{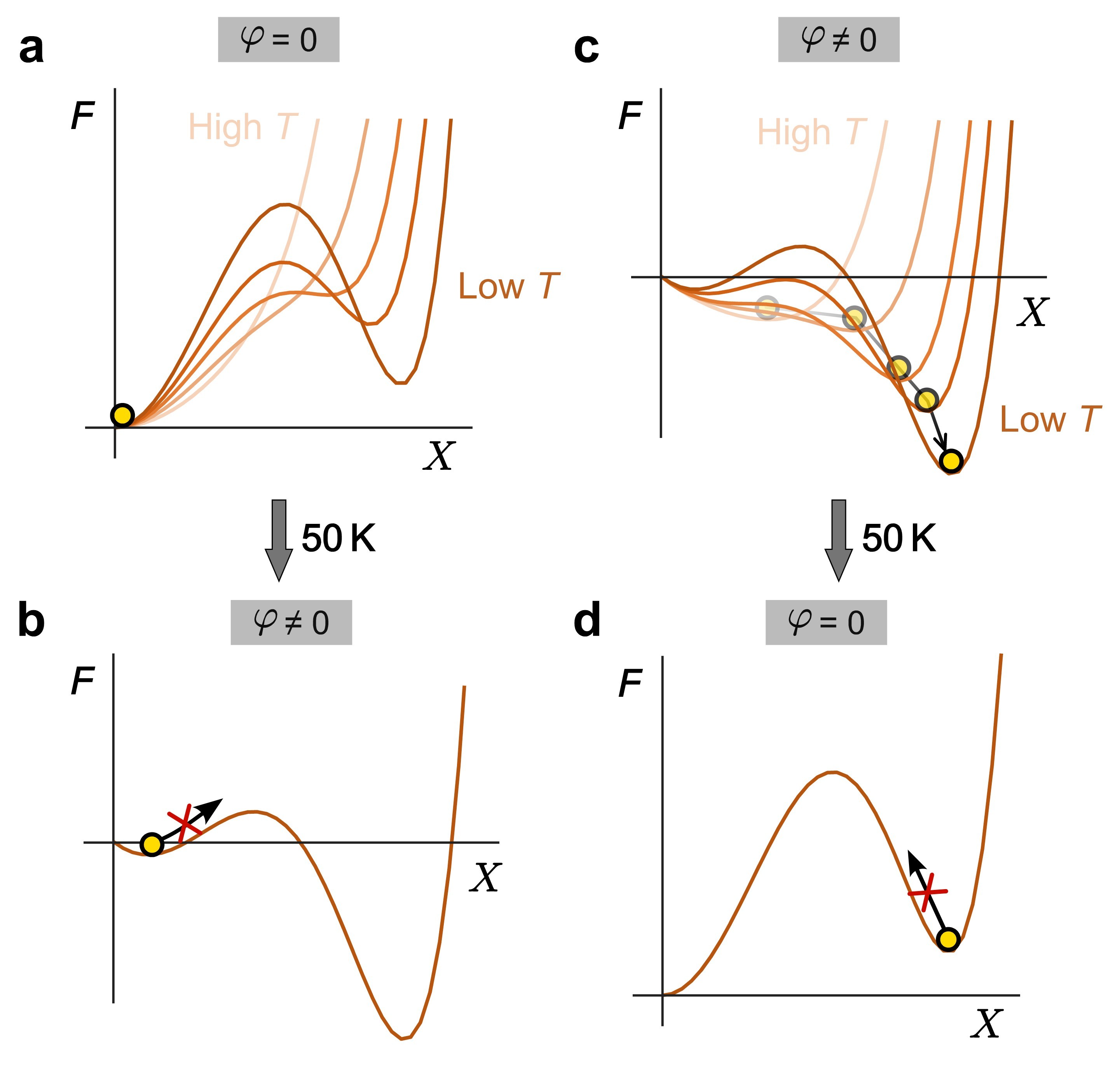}
\caption{\justifying\let\raggedright\justifying
\figtitle{Free-energy evolution of the lattice order $X$ during the ``doping + cooling'' process.}
\textbf{a–b,} Cooling at $n<n_e$, i.e., in the absence of $\varphi$, the free-energy landscape favors the $X=0$ minimum, and the system will naturally fall into the superlattice OFF state.  
\textbf{c–d,} Cooling at $n\geq n_e$, i.e., in the presence of $\varphi$, the coupling term $\varphi X$ tilts the free-energy landscape, making the $X\neq0$ state the global minimum. As a result, the system settles into the superlattice ON state.  
At low temperatures, the energy barrier separating the $X=0$ and $X\neq0$ minima increases and becomes sufficiently large that the coupling term $\varphi X$ is no longer strong enough to drive transitions between the two states. Consequently, if the system cools into the $X=0$ state, it remains trapped there even when $\varphi$ is subsequently applied (\textbf{b}); conversely, if the system cools into the $X\neq0$ state in the presence of $\varphi$, removing $\varphi$ does not restore the $X=0$ state, again due to the large energy barrier that must be overcome (\textbf{d}). In this low-temperature regime, the system is therefore effectively locked into whichever lattice state is selected during cooling.
}
\label{Free_energy_cooling_ONOFF}
\end{figure*}

\begin{figure*}
\includegraphics[width=\textwidth]{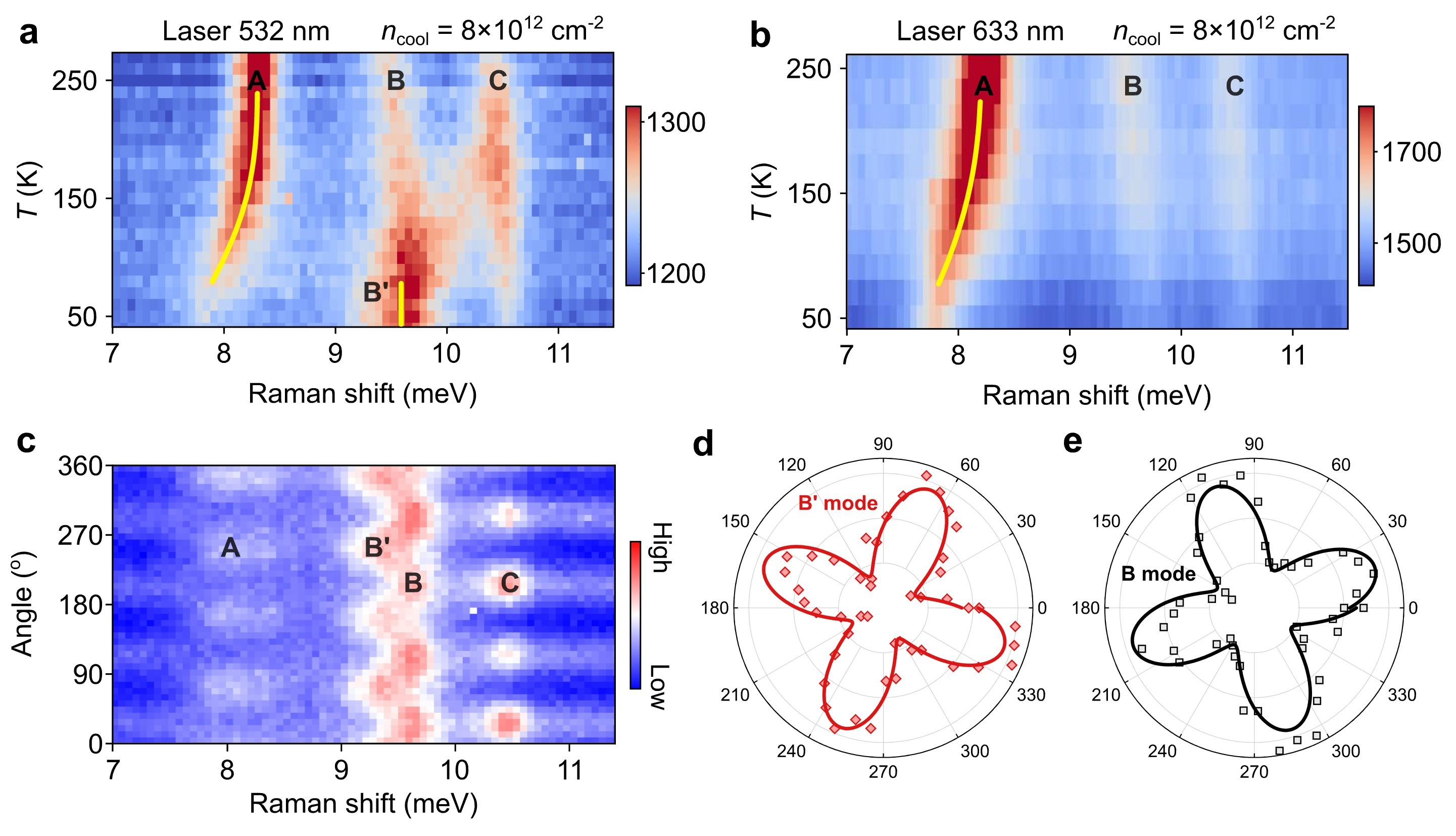}
\caption{\justifying\let\raggedright\justifying
\figtitle{Additional Raman evidence of B$^\prime$ mode.}
\textbf{a,} Raman intensity map measured in the left–right (L–R) circular polarization configuration using a 532~nm excitation laser during cooling from 270~K to 40~K at a fixed carrier density of $n_\mathrm{cool}=8\times10^{12}$~cm$^{-2}$.
\textbf{b,} Raman intensity map measured under identical cooling and doping conditions using a 633~nm excitation laser. In both cases, three Raman modes (A, B, and C) are observed at high temperature, with mode A softening upon cooling. Notably, mode B$^\prime$ is selectively enhanced under 532~nm excitation and is absent under 633~nm excitation, indicating a resonant enhancement effect.
\textbf{c,} Angle-dependent Raman spectra measured in the linear cross-polarization configuration (X–Y), where the angle of incident and scattered light polarization is controlled using half-wave plate, while keeping the polarizer and analyzer fixed in orthogonal position. Four distinct modes (A, B, B$^\prime$, and C) are clearly resolved.
\textbf{d--e,} Polarization-angle dependence of modes B$^\prime$ and B, revealing distinct symmetry behaviors and confirming that B and B$^\prime$ correspond to separate Raman modes. Additional Raman data and discussions are provided in SI Section 8. 
}
\label{Raman_red_green}
\end{figure*}

\end{document}